\documentclass[useAMS,usenatbib]{mn2e}

\usepackage{graphics}
\usepackage{epsfig}
\usepackage{graphicx}

\title[Assembly Histories]{The SLUGGS Survey:
The Assembly Histories of Individual Early-type Galaxies
}
\author[D. A. Forbes et al.]{Duncan A. Forbes$^{1}$\thanks{E-mail:
dforbes@swin.edu.au}, Aaron J. Romanowsky$^{2,3}$, Nicola Pastorello$^{1}$, Caroline Foster$^{4}$, 
\newauthor
Jean P. Brodie$^{3}$, Jay Strader$^{5}$, Christopher Usher$^{6}$, Vincenzo Pota$^{3}$ 
\\
$^{1}$Centre for Astrophysics \& Supercomputing, Swinburne University, Hawthorn VIC 3122, Australia\\
$^{2}$Department of Physics and Astronomy, San Jos\'e State
University, One Washington Square, San Jose, CA 95192, USA\\
$^{3}$University of California Observatories, 1156 High Street,
Santa Cruz, CA 95064, USA\\
$^{4}$Australian Astronomical Observatory, PO Box 915, North Ryde, NSW 1670, Australia\\
$^{5}$Department of Physics and Astronomy, Michigan State
University, East Lansing, Michigan 48824, USA\\
}
\begin{document}


\pagerange{\pageref{firstpage}--\pageref{lastpage}} \pubyear{2002}

\maketitle

\label{firstpage}

\begin{abstract}

Early-type (E and S0) galaxies may have assembled via a variety of
different evolutionary pathways. Here we investigate these pathways by
comparing the stellar kinematic properties of 24 early-type galaxies
from the SLUGGS survey with the hydrodynamical simulations of Naab et
al. (2014). In particular, we use the kinematics of starlight up to 4
effective radii (R$_e$) as diagnostics of galaxy inner and outer
regions, and assign each galaxy to one of six Naab et al. assembly classes.

The majority of our galaxies (14/24) have kinematic characteristics
that indicate an assembly history dominated by gradual gas
dissipation and accretion of many gas-rich minor mergers. Three 
galaxies, all S0s, indicate that they have experienced gas-rich major
mergers in their more recent past.  One additional elliptical galaxy is
tentatively associated with a gas-rich merger which results in a
remnant galaxy with low angular momentum.  Pathways dominated by
gas-poor (major or minor) mergers dominate the mass growth of six 
galaxies.  Most SLUGGS galaxies appear to have grown in mass (and size) via the
accretion of stars and gas from minor mergers, with late major mergers
playing a much smaller role.

We find that the fraction of accreted stars correlates with the
stellar mean age and metallicity gradient, but not with the slope of
the total mass density profile. We briefly mention future
observational and modelling approaches that will enhance our ability
to accurately reconstruct the assembly histories of individual present
day galaxies.

\end{abstract}


\section{Introduction}

Galactic archaeology is the detailed study of nearby galaxies that aims to 
reconstruct their evolutionary histories. A key, but often elusive, goal is 
to understand the assembly of an individual galaxy in terms of 
its initial formation and subsequent mass growth 
via mergers/accretion of stars and gas. 

The formation of massive galaxies (which are
predominately early-type) is often described in two stages or phases.
In the first phase, it is thought that gas collapses via strong radial
inflows, into a turbulent, clumpy rotating disc (Keres et al. 2005;
Dekel et al. 2009; Ceverino et al. 2010; Zolotov et al. 2015). The vast bulk of the stars
form in-situ within the central galaxy regions.  In the second phase,
mass growth occurs via stars that are formed ex-situ, i.e. within
other galaxies that are later accreted/merge onto the host galaxy (Naab et
al. 2007; Oser et al. 2010, 2012; Lackner et al. 2012; Dubois et al. 2013). 
In this picture, 
accretion may dominate the mass 
growth at low redshift, with more massive galaxies accreting a 
larger fraction of their mass (Khochfar et al. 2011). 

Recently, Naab et al. (2014) presented detailed cosmological zoom
simulations with hydrodynamics of 44 massive galaxies formed in
two-phases as described above, while also placing them in the context of 
previous galaxy formation modelling.  
In particular, they modelled the shapes
and kinematics of their present day simulated galaxies in two
dimensions (2D) with a spatial resolution of 0.4 kpc (about 1/10 of the 
typical effective radius). This included 2D kinematics of velocity, velocity
dispersion ($\sigma$) and higher order moments h$_3$ (which indicates the 
skewness of the velocity distribution) and
h$_4$ (which indicates the kurtosis, or peakiness, of the velocity 
distribution) similar to that provided by today's integral field
spectroscopy observations.  Different mass assembly histories leave
distinct imprints in the 2D kinematic maps and 1D kinematic 
profiles of the final galaxy. From
these imprints 6 distinct evolutionary
pathways were identified. These pathways involved different merger ratio contributions 
(minor or
major), timescales (early or recent mergers) and gas contents (rich or
poor). Although only a small sample, these represent possible pathways 
for the growth of massive galaxies in a cosmological context.

Using the same suite of models as Naab et al. (2014), 
Wu et al. (2014) showed that these
different assembly pathways were also reflected in the outer kinematic 
radial profiles of present day simulated galaxies.
The different assembly histories are also naturally associated with different
mean fractions of stars formed 
in-situ and ex-situ (i.e. accreted). In particular, the
fraction of stars formed ex-situ is predicted to correlate with galaxy
stellar mass and mean stellar age (Naab et al. 2014).  
Using similar initial conditions to 
the Naab et al. (2014) models, Remus et al. (2013) and Hirschmann
et al. (2015) showed that total mass density profiles and stellar
metallicity outer gradients respectively, also correlate with the ex-situ
fraction.

Using a semi analytic model approach, Khochfar et al. (2011) focused on 
the assembly history of centrally fast and slow
rotators in a similar cosmology
to Naab et al. (2014).  
They concluded that slow rotators accreted over 50-90\% of their mass
in 1-2 major mergers and a series of minor mergers. For fast rotators
the accreted fraction is less than 50\% and is dominated by minor
mergers. This picture agrees in broad strokes with that of Naab et
al. (2014), although the hydrodynamic simulations suggest that a fraction of 
slow rotators may have experienced only minor mergers. Moody et al. (2014) also 
focused on reproducing fast and slow rotators. In their controlled 
binary and multiple merger simulations, they found that binary mergers almost always 
result in a fast rotator while round, slow rotators as observed in nature require multiple 
mergers. 

The first large 2D kinematic survey of early-type galaxies was 
the ATLAS$^{\rm 3D}$ survey (Cappellari et al. 2011). They found a
variety of kinematic structures, including the prevalence of embedded kinematic
discs.  Based on this, they classified 
86\% of their 260 early-type galaxies to be centrally fast or regular 
rotators (i.e. disc-like kinematics with high angular momentum in the inner 
regions) and 12\% to be slow
rotators with low angular momentum. From the SLUGGS 
(SAGES Legacy Unifying Globulars and GalaxieS, http://sluggs.swin.edu.au) survey, 
Arnold et al. (2014) and Foster et
al. (2015) found that centrally slow rotators maintain their slow
rotation at larger radii, but centrally fast rotators show significant 
diversity. Their rotation may plateau, decline or steepen with increasing 
radius. 
Similar results were obtained from observations of PNe
(Coccato et al. 2009; Cortesi et al. 2013).  
As well as SLUGGS, a number of other 
2D integral field spectroscopy surveys are underway or near completion
e.g. CALIFA, MASSIVE, SAMI, MaNGA (see Brodie et
al. 2014 for a summary). Each has its own comparative advantages.

A number of previous studies have compared detailed observations with
tailored models to infer the past history of individual
galaxies. However, we are aware of only one independent study (Spiniello et al. 2015)
that has directly compared their own 2D results with the simulations of
Naab et al. (2014) and thus assigned an assembly class within the
two-phase model of galaxy formation.  The Spiniello et al. (2015) data
reached out to 0.7 effective radii (R$_e$) in NGC 4697 
and confirmed the disc-like kinematics found
by ATLAS$^{\rm 3D}$ and the SLUGGS surveys (Krajnovic et al. 2011; Arnold et
al. 2014). They found its kinematic properties to be most consistent
with class A, i.e. having undergone gas-rich minor mergers with
gradual dissipation (however a recent gas-rich major merger may also
be a possibility). From their stellar population analysis, they determined that the disk formed some 3 Gyr after the bulge. 

Here we use the high S/N, large radial extent and excellent velocity
resolution of the 2D kinematics from the SLUGGS survey of 
nearby
early-type galaxies (Brodie et al. 2014) to compare their stellar properties
with those of the simulated galaxies of Naab et al. (2014). Combining
with the central ($<$1 R$_e$) kinematics from the ATLAS$^{\rm 3D}$ survey where
available, we classify our galaxies into one of the six classes of
assembly history. We then compare their V/$\sigma$ profiles with
average values from the simulated galaxies of Wu et al. (2014), and
the average ex-situ mass fractions of different assembly classes with
mean stellar age, mass density profile slope and stellar metallicity
gradients. Finally, we discuss the relative frequency of the different assembly classes 
for the SLUGGS galaxies and contrast them with model predictions.

\section{Assembly class characteristics}

Based on cosmological zoom simulations of 44 model galaxies, Naab et
al. (2014; hereafter N14) defined 6 `assembly classes' with distinct
evolutionary pathways. The galaxies build up their mass via mergers
and accretions, at different times, with variations in gas
content. This gives rise to a range in the relative fraction of
ex-situ formed stars to the total stellar mass. The present day total
stellar masses of their galaxies span 
2.6 $<$ M$_{\ast}$ $<$ 57 $\times$
10$^{10}$ M$_{\odot}$, and lie within dark matter halos of mass 
2.2 $<$ M$_{\ast}$ $<$ 370 $\times$
10$^{11}$ M$_{\odot}$.
The different pathways are 
encoded in the {\it
  present day} isophotal shapes and 2D kinematics of the
galaxies. We note that the resolution and feedback of the models are limited and refer the reader to N14 for a 
see a discussion.

Below we briefly describe the key properties of each N14 assembly
class. We focus on 2D kinematics since  the isophotal shape of galaxies
does not provide much leverage between the different classes (nor do
the models cover the full range of observed ellipticities). The 2D
kinematic properties are based on the appearance of the 2D velocity
and velocity dispersion maps out to 2 R$_e$, and the distribution of
individual pixel values of the higher order moments h$_3$ and h$_4$ versus
V/$\sigma$ within 1~R$_e$. A 1D property is also included by N14,
i.e. the shape of the $\lambda_R$ profile (a proxy for projected
angular momentum) to 2~R$_e$ as defined by Emsellem et al. (2004). \\

\begin{table*}
\caption{Naab et al. (2014) assembly class properties}
\begin{tabular}{lcccccccc}
\hline
Class & Class & Central & Gas  & M$_{ins}$/M$_{\ast}$ & $\epsilon$ & $\lambda_R$ & 2D map & h$_3$, h$_4$\\
      & fraction & rotation &  rich/poor      &     mean               &  range     &  profile & feature & trend\\
 (1) & (2) & (3) & (4) & (5) & (6) & (7) & (8) & (9) \\
\hline
A & 21\% & Fast & Rich & 0.27 & 0.3--0.55 & peaked & $\sigma$ dumbell 
& anti,V-shape \\
B & 16\% & Fast & Rich & 0.29 & 0.3--0.55 & rising & V disc & anti, 0\\
C & 16\% & Slow & Rich & 0.24 & 0.2-0.4 & flat & $\sigma$ dip & 0,0\\
D & 11\% & Fast & Poor & 0.11 & 0.3--0.5 & rising & featureless & 0,0\\
E & 25\% & Slow & Poor & 0.14 & 0.25--0.6 & rising & KDCs/twists & range, range\\
F & 11\% & Slow & Poor & 0.13 & 0.15--0.4 & flat & featureless & range, range\\
\hline

\end{tabular}
\\
Notes: (1) Assembly class (2) Fraction of 44 simulated model galaxies; (3) Central rotator type; (4) Gas-rich or gas-poor merger history; (5) Mean in-situ to  total stellar mass fraction since z = 2; (6) Range of edge-on 
ellipticity values at R$_e$; (7) Shape of $\lambda_R$ profile to 
2R$_e$; 
(8) Main 2D kinematic map feature; (9) h$_3$ and h$_4$ vs 
V/$\sigma$  behaviour (anti = anti-correlation, V-shape = V shaped distribution, 0 = h$_3$ or h$_4$ clustered around zero, range = large range in h$_3$ or h$_4$ values with V/$\sigma$ clustered around zero.). 
\end{table*}

\noindent
{\bf Class A}\\
Centrally fast, discy rotators that have undergone gas-rich minor ($<$ 1:4) mergers 
and no major ($>$ 1:4) mergers in the last $\sim$8 Gyr. They reveal 
a strong h$_3$ vs V/$\sigma$ anti-correlation and 
a V-shaped h$_4$ vs V/$\sigma$ distribution.
The $\lambda_R$ profile may show a peak and then 
fall by 2~R$_e$. Their 2D kinematic maps 
often show a distinctive `dumbbell-like'
velocity dispersion 
enhancement along the minor axes (i.e. perpendicular to the disc). \\

\noindent
{\bf Class B}\\
Fast, discy rotators that have experienced a gas-rich major merger since 
z = 2. Like class A, they show 
a strong h$_3$ vs V/$\sigma$ anti-correlation but the h$_4$ values are around zero (i.e. h$_4$ is largely independent of V/$\sigma$).
Unlike class A, the $\lambda_R$ profiles continue to rise beyond 1~R$_e$. Their 2D kinematic maps
show a slight velocity dispersion dip at the centre but no strong enhancement along the minor axes. These galaxies 
have the highest mean in-situ fraction in the last 10 Gyr 
(and are therefore expected to have 
steep total mass density profiles and 
steep negative metallicity gradients; see Section 6). \\

\noindent
{\bf Class C}\\
Centrally slow rotators that have experienced a recent gas-rich major merger. There is little, or no, 
correlation between h$_3$ or h$_4$ and V/$\sigma$. The $\lambda_R$ profiles are flat. Their 2D kinematic maps show 
a distinctive central velocity dispersion dip. \\

\noindent
{\bf Class D}\\
Fast rotators that have experienced a recent gas-poor major merger. Both h$_3$ and h$_4$ vs V/$\sigma$ distributions are clustered 
around zero. They reveal a rising $\lambda_R$ profile, with largely featureless kinematic maps. These galaxies have the lowest
in-situ fraction in the last 10 Gyr (and are 
therefore expected to have shallow total mass density profiles 
and shallow metallicity gradients). \\

\noindent
{\bf Class E}\\
Centrally slow rotators that have experienced a recent gas-poor major merger. Both h$_3$ and h$_4$ 
show a very weak anti-correlation and a large range of values, when compared to V/$\sigma$.  
They show slowly rising $\lambda_R$ profiles. Their 2D kinematic maps sometimes show kinematically distinct cores (KDCs) or kinematic twists. 
They tend to be quite elongated galaxies with a mean ellipticity of 0.43. \\

\noindent
{\bf Class F}\\
Slow rotators that have undergone a gas-poor minor merger. Like class E, they show a very weak 
anti-correlation of h$_3$ and h$_4$ with V/$\sigma$, and a large range of h$_3$ and h$_4$ values. They have low angular momentum with 
flat $\lambda_R$ profiles. Their 2D kinematic maps are largely featureless. They tend to be near-round galaxies with a mean ellipticity of 0.27. \\

The properties of the six N14 assembly classes are summarised in Table
1. Galaxies of classes D, E and F have experienced higher fractions of
accreted stars compared to classes A, B and C. The
former galaxies tend to have shallower metallicity gradients
(Hirschmann et al. 2015), shallower total mass density profiles (Remus
et al. 2013), and to be more massive on average.  The fast rotator
classes (i.e. A, B and D) have V/$\sigma$ values at 5~R$_e$ similar to
those at 1 R$_e$, whereas the slow rotators (C, E and F) have increased 
rotational support by 5 R$_e$ (Wu et al. 2014).

\section{The SLUGGS sample}

The SLUGGS survey (Brodie et al. 2014) investigates the spatial,
kinematic and chemical properties of 25 nearby early-type
galaxies. Using the DEIMOS spectrograph on the Keck telescope we
obtain kinematics and metallicity information for the galaxy starlight
out to $\sim$3 R$_e$ and for globular clusters out to $\sim$8 R$_e$.
Although only a small sample, it is chosen to be representative of the
early-type galaxy population.  Data collection for the SLUGGS survey
is still underway in 2015, nevertheless we have obtained results for
the bulk of the sample. Here we draw on the stellar kinematic measurements of
Arnold et al. (2014) and 
Foster et al. (2015), stellar metallicity gradients from Pastorello et
al. (2014) with updates (see Appendix B) and total mass density
profile slopes from Cappellari et al. (2015).

\begin{table}
\caption{SLUGGS galaxy properties}
\begin{tabular}{lccccc}
\hline
NGC & M$_{\ast}$ & Type & $\epsilon$ & $\gamma_{tot}$ & Age \\
          &  (10$^{10}$ M$_{\odot}$) & & & & (Gyr)\\
 (1) & (2) & (3) & (4) & (5) & (6)\\
\hline
720 & 22.4 & E5 & 0.49 & --& --\\
821 & 9.3 & E6 & 0.35 & -2.23 & 12.9\\
1023 & 9.5 & S0 & 0.63 & -2.20 & 13.5\\
1400 & 13.3 & E1/S0 & 0.13 & -- & 13.8\\
1407 & 39.6 & E0 & 0.07 & -- & 12.0\\
2768 & 18.9 & E6/S0 & 0.57 & -2.01 & 13.3\\
2974 & 6.5 & E4/S0 & 0.37 & -2.30 & 11.8\\
3115 & 9.4 & S0 & 0.66 & -2.37 & 9.0\\
3377 & 2.8 & E5-6 & 0.33 & -2.05 & 11.3\\
3608 & 6.7 & E1-2 & 0.20 & -- & 13.0\\
4111 & 4.6 & S0 & 0.79 & -2.13 & 6.0\\
4278 & 7.6 & E1-2 & 0.09 & -2.29 & 13.7\\
4365 & 30.4 & E3 & 0.24 & -- & 13.4\\
4374 & 28.6 & E1 & 0.05 & -- & 13.7\\
4473 & 7.4 & E5 & 0.43 & -2.18 & 13.0\\
4494 & 10.5 & E1-2 & 0.14 & -2.26 & 11.0\\
4526 & 17.1 & S0 & 0.76 & -2.24 & 13.6\\
4564 & 3.8 & E6 & 0.53 & -- & 13.3\\
4649 & 35.9  & E2/S0 & 0.16 & -2.19 & 13.2\\
4697 & 10.7 & E6 & 0.32 & -2.23 & 13.4\\
5846 & 25.2 & E0-1/S0 & 0.08 & -- &12.7\\
7457 & 1.9 & S0 & 0.47 & -2.23 & 6.1\\
\hline
3607 & 19.7 & S0 & 0.13 & --& 13.5\\
5866 & 9.4 & S0 & 0.58 & --& 8.4\\
\hline

\end{tabular}
\\
Notes: (1) NGC name; (2) Total stellar mass (see text for details); (3) Morphological type; (4) Ellipticity for the outer isophotes from Brodie et al. (2014); (5) Total mass density slope $\gamma_{tot}$ from Cappellari et al. (2015); (6) Mass-weighted stellar age within 1 R$_e$ 
from McDermid et al. (2015), Norris et al. (2006) for NGC 3115 
and Spolaor et al. (2008) for NGC 1400 and NGC 1407. The last two galaxies in 
the Table are SLUGGS bonus galaxies.\\
\end{table}

The properties of SLUGGS galaxies (and a few bonus galaxies observed in addition to the main survey) are summarised in Brodie et al. (2014). Here, in 
Table 2 we list some properties relevant for this work. In particular, we calculate the total
stellar mass of each galaxy starting with the K-band luminosities
listed in Brodie et al. (2014). This is corrected for missing flux using
the formula given by Scott et al. (2013) and we apply M/L$_K$ = 1
(i.e. appropriate for old stellar populations with a Kroupa 
IMF). 
The range in stellar mass of SLUGGS galaxies 
is well matched to the 44 simulated galaxies of N14. 
We also list the total mass density slopes from 0.1 to 4 R$_e$ for 
the combined fitting 
of ATLAS$^{\rm 3D}$ data in the central regions and SLUGGS data further out
for 14 galaxies (Cappellari et al. 2015). Stellar metallicity gradients to 2.5~R$_e$, first
presented in Pastorello et al. (2014), have been remeasured and are given in 
the Appendix B. 

Central stellar ages for most SLUGGS galaxies are available from the ATLAS$^{\rm 3D}$ work of
McDermid et al. (2015).  Here we use the mass-weighted age within 1
R$_e$. For NGC 1400 and NGC 1407, we use the luminosity-weighted ages
from Spolaor et al. (2008), which reveal a constant age to 1 R$_e$
(i.e. half of the galaxy stellar mass). For NGC 3115 we use the
average of the spheroid and disk age out to 1 R$_e$ from Norris et
al. (2006).

\section{Assembly classes of SLUGGS galaxies}

\subsection{h$_3$ and h$_4$ versus V/$\sigma$}

In Appendix A we show plots of the behaviour of the higher order velocity
moments h$_3$ and h$_4$ against V/$\sigma$ for individual locations 
within 
each galaxy. The data, from the kinematic fitting of Foster et
al. (2015), have been restricted to those points having S/N $>$ 10
(which corresponds to an uncertainty in h$_3$ and h$_4$ of about
$\pm$0.15 and $\pm$0.1 in V/$\sigma$). 
These Figures can be compared to the N14 models shown in
their figure 9, although those only showed the distribution of points
within 1 R$_e$.  Our plots show the values within, and beyond, 1
R$_e$ using different symbols, where R$_e$ is taken from Brodie et al. 
(2014). 

The distribution of h$_3$ and h$_4$ versus V/$\sigma$ at individual
positions in a galaxy provide important diagnostics. The h$_3$ term
measures the skewness of the line-of-sight velocity distribution
(LOSVD), and h$_4$ measures the kurtosis or peakedness of the
LOSVD. The V/$\sigma$ ratio is a good proxy for the ratio of ordered
rotation to random motions. For example, an LOSVD with steep leading
wings, indicative of a rotating disc, shows an anti-correlation of h$_3$
versus V/$\sigma$.

In general, the datapoint trends within, and beyond, 1R$_e$ are very
similar, with the $>$ 1 R$_e$ data points showing a larger range in
values (which may be due to increased observational error). 
However, some exceptions
exist. For example, for NGC 2974 the h$_4$ values within 1 R$_e$ have a
bias to positive h$_4$ values, whereas beyond 1 R$_e$ the data scatter
more evenly about zero. For NGC 3377, the h$_4$ values are close to zero, 
indicative
of disc-like kinematics, but scatter to larger/smaller h$_4$ values beyond
1 R$_e$. This behaviour is consistent with the 2D kinematic map which
reveals an inner fast rotating disc surrounded by a slow
pressure-supported halo (see Arnold et al. 2014; Foster et al. 2015
for details).

The h$_3$ and h$_4$ distributions show a variety of patterns similar
to those shown in figure 9 of N14, e.g. the anti-correlated h$_3$ and
V/$\sigma$ characteristic of class A (e.g. NGC 3377), the V-shaped
h$_4$ distribution of class B (e.g. NGC 4526), the large range in
h$_3$ and h$_4$ and small range in V/$\sigma$ of classes C, E and F
(e.g. NGC 5846). Class D, with a small range in both h$_3$ and h$_4$, is tentatively assigned to only one galaxy (NGC 4649). 
We find some galaxies to have `flat' distributions,
i.e. both h$_3$ and h$_4$ have values close to zero while there is
a large range in V/$\sigma$ (e.g. NGC 7457).  This behaviour is not
clearly seen in the N14 models, however it may be related to classes A and B
(due to a very weak anti-correlation of h$_3$ vs V/$\sigma$).

On the basis of a qualitative comparison with figure 9 of N14 we
assign assembly classes, and list them in Table 3.  
We could not classify the h$_3$ and h$_4$
distribution of one galaxy (NGC 1400). In some cases, one assembly
class clearly best resembles our data; in others several assembly
classes may be compatible.

\begin{table}
\caption{SLUGGS galaxies assembly classes}
\begin{tabular}{lcccc}
\hline
NGC & h$_3$, h$_4$ & $\lambda_R$ & 2D map & Class\\
 (1) & (2) & (3) & (4) & (5) \\
\hline
720  & EF & BDEF & ABDE & A? (E)\\
821  & ACEF & A & ABDE & A (E)\\
1023 & AB & B & BD & B (A)\\
1400 & -- & ABDE & AE & A (E)\\
1407 & CEF & BD & CD & C? (D)\\ 
2768 & AB & BD & ABCD & B (A)\\
2974 & AB & AB & AB & A (B)\\
3115 & AB & AB & AB & A (B)\\
3377 & A & A & AB & A \\
3608 & CDEF & BDE & E & E (D)\\
4111 & AB & A & AB & A (B)\\
4278 & CDEF & A & ABE & A (E)\\
4365 & CDEF & DE & E & E (D) \\ 
4374 & CEF & CEF & EF & F (E)\\
4473 & CEF & ACEF & AE & E (A)\\
4494 & ABD & AB & AC & A (B)\\
4526 & B & B & AB & B (A)\\
4564 & AB & A & AB & A (B)\\
4649 & ABD & BD & BCD & D? (B)\\
4697 & A & A & A & A\\
5846 & CEF & CF & CEF & F? (C)\\
7457 & B & AB & AB & A? (B)\\
\hline
3607 & CD & CF & ABCE & A? (C)\\
5866 & AB & BD & AC & A (B)\\
\hline

\end{tabular}
\\
Notes: (1) NGC name; (2)  h$_3$ and h$_4$ vs V/$\sigma$  behaviour; 
(3) Shape of $\lambda_R$ profile; 
 (4) 2D kinematic map features; (5) Final assembly class assignment (see Section 4.3 for details). 
\end{table}

\subsection{2D kinematic maps and 1D $\lambda_R$ profiles}

2D kinematic maps of velocity, velocity dispersion, h$_3$ and h$_4$
have been published in Arnold et al. (2014) and Foster et al. (2015),
along with local $\lambda_R$ profiles. In Figures 1 and 2 we show an example of a 
2D velocity map and 1D $\lambda_R$ profile for NGC 1023.
The 2D maps and 1D $\lambda_R$
profiles typically reach at least 2 R$_e$ and in some cases 4 R$_e$
and beyond. We also use the 2D maps of the ATLAS$^{\rm 3D}$ survey which has
superior sampling in the central regions of galaxies (and use
Cappellari et al. 2007 for NGC 720 which is not part of the ATLAS$^{\rm 3D}$
survey but was observed with the SAURON instrument).  We note that the
$\lambda_R$ profiles published in the ATLAS$^{\rm 3D}$ survey are cumulative,
i.e. they are luminosity-weighted and so the profiles are dominated by
the central region properties. To probe the radial angular momentum in
the SLUGGS galaxies we focus on our own local $\lambda_R$ profiles.

\begin{figure}
\begin{center}
\includegraphics[scale=0.33,angle=0]{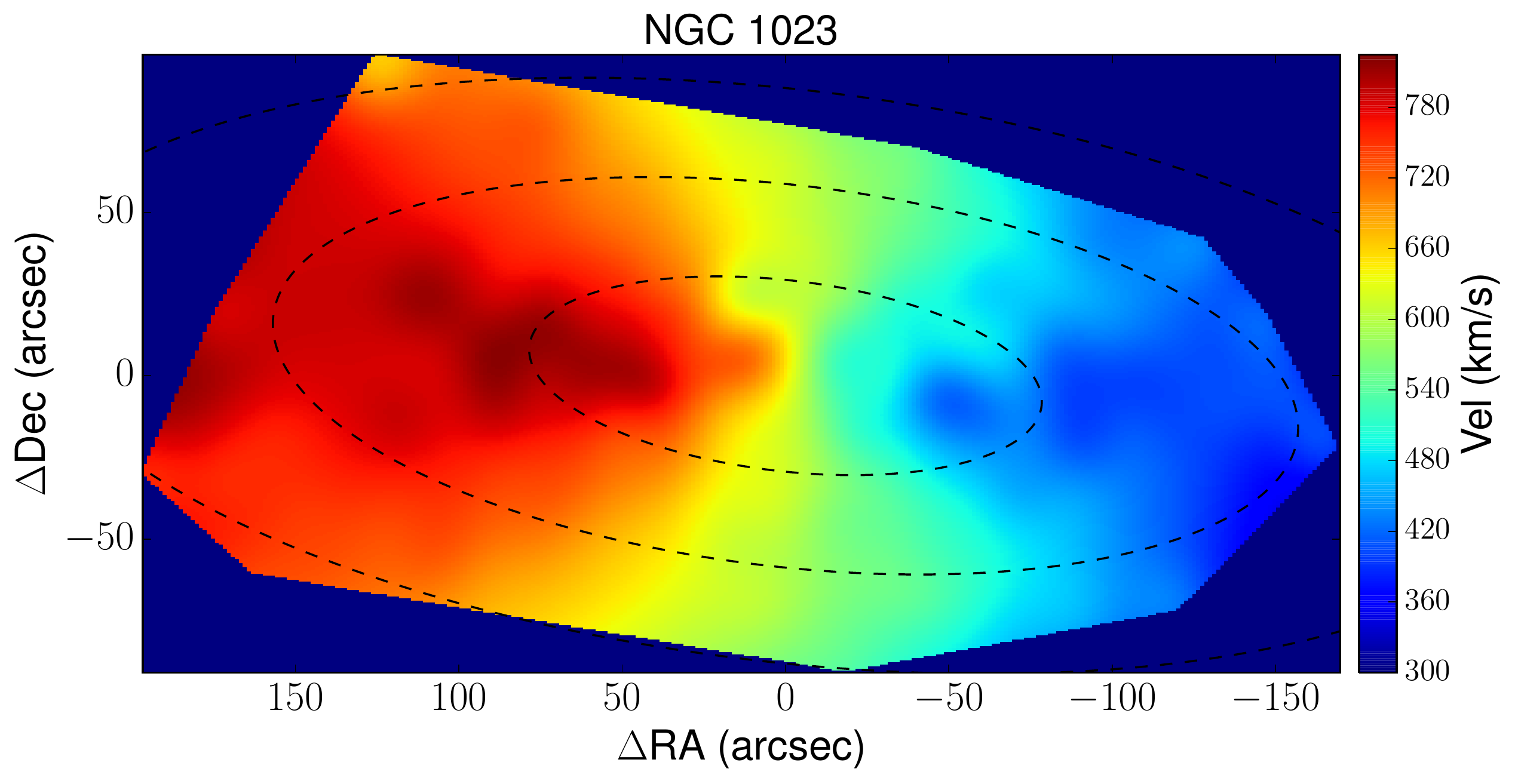}
\caption{2D stellar velocity map of NGC 1023 from SLUGGS data. 
The map has been interpolated 
in 2D based on the velocities measured at individual slit 
locations. Dashed lines show 
isophotes corresponding to 1, 2 and 3 R$_e$. A color bar
 indicates the rotation velocity.
}
\end{center}
\end{figure}

\begin{figure}
\begin{center}
\includegraphics[scale=0.5,angle=0]{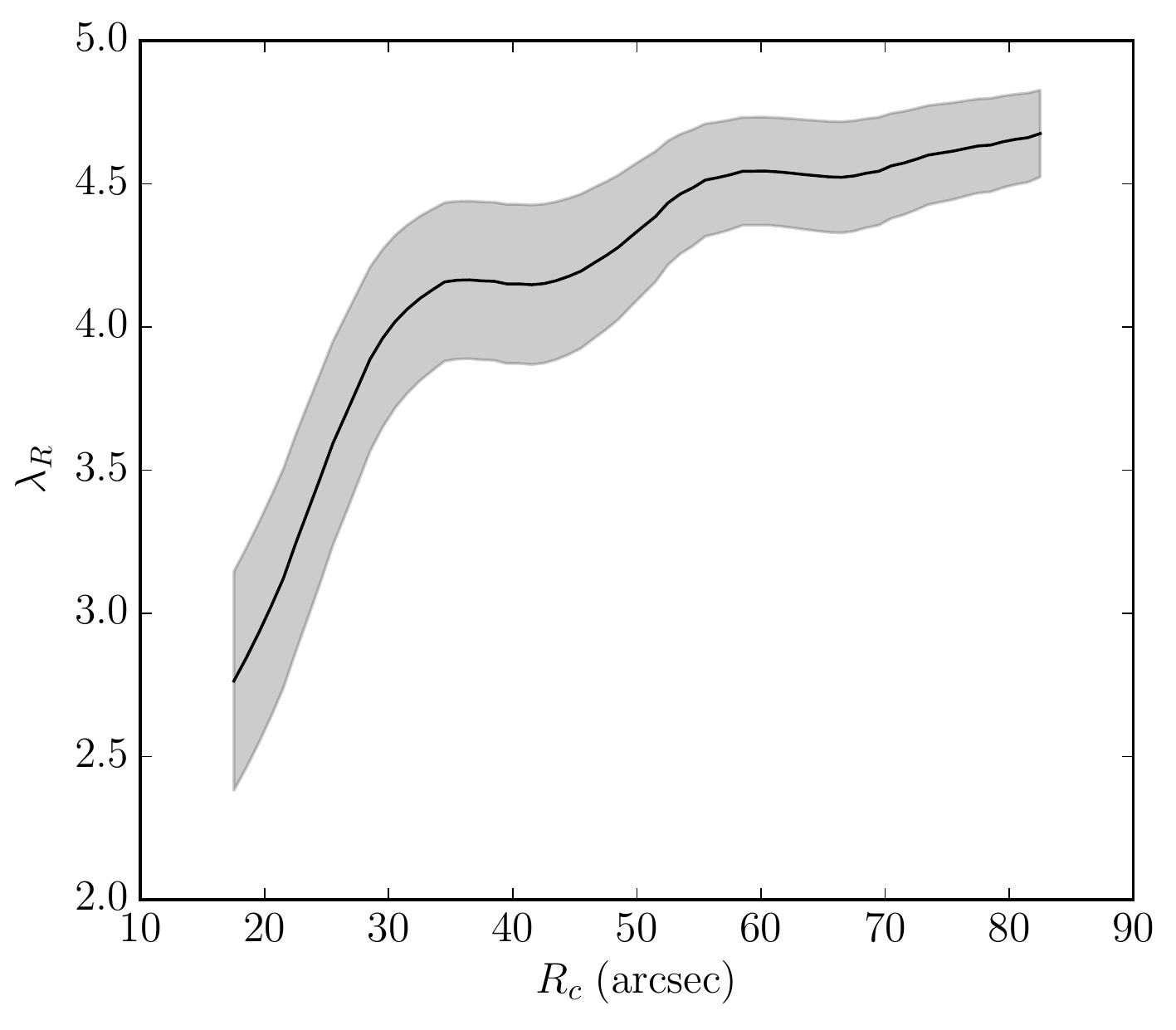}
\caption{Local $\lambda_R$ profile of NGC 1023 from SLUGGS data. 
The solid line shows the profile as a function of circularised radius. 
The shaded region shows the 1$\sigma$ 
uncertainty envelope. 
}
\end{center}
\end{figure}

After visually examining the 2D maps and the
1D profiles we assign assembly classes based on their
characteristics as described by N14. As with the h$_3$ and h$_4$ distributions, 
2D maps and $\lambda_R$ profiles are often consistent with 
more than one assembly 
class and we list them all in Table 3.


The PNe study of Coccato et al. (2009) included nine SLUGGS
galaxies. In particular, they examined the cumulative 
1D $\lambda_R$ profiles to
$\sim$7 R$_e$ based on PNe kinematics.  Most galaxies show good agreement with our stellar
light local $\lambda_R$ profiles, including the peaked and then falling
profile of NGC 3377 (Foster et al. 2015). However, a couple of
exceptions exist. For NGC 4494 our $\lambda_R$ profile continues to
rise out to 3 R$_e$ whereas the PNe-based profile falls by 1
R$_e$. Differences could be due to the complex halo kinematics of NGC
4494 (Foster et al. 2011). In the case of NGC 1023, the profiles agree
to 3 R$_e$ (i.e. the outermost starlight point) but beyond that the
PNe indicate a falling $\lambda$ profile.

\subsection{Final assignment of assembly class}

We have qualitatively assigned assembly classes to each SLUGGS galaxy
based on its 1D local $\lambda_R$ profile, h$_3$ and h$_4$ versus
V/$\sigma$ behaviour and 2D kinematic map features, 
independently. These are listed in columns 2, 3 and 4 of Table 3.  
There is some variation in the assigned class(es) depending on the diagnostic used. This is 
to be expected as some diagnostics do not discriminate well between different assembly classes.
For example, the h$_3$ and h$_4$ distributions of class E and F are almost identical. 
In assigning the final assembly class we have given more weight to the 2D map than to the 
other diagnostics. In
column 5 of Table 3 we list the class that best represents the galaxy. In a few
cases, the class assignment is tentative (denoted by a ?) and for most
galaxies we give an alternative class (denoted in parentheses). An
important caveat when comparing observations to simulations is that
the observed galaxies have a range of (largely random) inclinations,
while simulations can be projected along chosen sight lines. An
illustration of inclination effects on cumulative $\lambda_R$ profiles
is shown by Wu et al. (2014) in their figure 12. They showed that
face-on profiles have less diversity and overall lower absolute values
than galaxies viewed edge-on. Inclination effects will be at least as strong 
for local $\lambda_R$ and V/$\sigma$ profiles that we use in this study.

We find that several galaxies reveal inner disc-like kinematics on varying 
scales of effective radii. These discs then give way to a more 
slowly rotating outer region. Following N14, we assign these galaxies  
to class A. 
We note that NGC 720 was observed 
by Cappellari et al. (2007), who found it to have
kinematics borderline between fast and slow rotators. We find it shows 
weak central 
disc-like kinematics and give it a tentative assignment of class A?.
We assign 14 galaxies to class A or A?, 3 galaxies to class B, 1 to C?, 
1 to D?, 3 to class E, and 2 to class F or F?. 

Most of the alternatives listed for class A are class B, 
which is also a fast disky rotator but with  an increased role for major, rather 
than minor, mergers. Five galaxies have class E listed as their alternative, compared to 
only 3 as the primary choice. If these E class alternatives are correct, it suggests that gas-poor major mergers 
are the second most frequent assembly pathway for our sample.

\section{V/$\sigma$ profiles}

Wu et al. (2014) examined the kinematic profiles for 42 model galaxies
from the same suite of zoom simulations as N14. As well as showing
cumulative $\lambda_R$ and V$_{rms}$ profiles, they plotted local
V/$\sigma$ profiles for edge-on model galaxies 
and tabulated their values at 5 R$_e$ relative to 1
R$_e$ (values for V$_{rms}$ were not given).

Using data from Foster et al. (2015) we plot the local V/$\sigma$
profiles for the SLUGGS galaxies, normalised at 1~R$_e$. In Figure 3
we show the galaxies that we have classified as slow rotators (classes
C, E and F).  The slow rotators all have
rising profiles indicating relatively more rotational support at larger radii. 
Although they only reach to $\le$ 3 R$_e$, the
profiles appear consistent with the average values of V/$\sigma$ at
5~R$_e$/1~R$_e$ for the 18 zoom models with slow rotator assembly classes
in Wu et al. (2014).

Galaxies classified as class A (i.e. disc-like kinematics as the result of
gas-rich minor mergers) reveal relatively gently rising profiles
(Fig. 4). This can be understood since several minor mergers will
tend to randomise and reduce the net velocities of halo stars (Di
Matteo et al. 2009).  In the case of NGC 3377 and NGC 4278, the fast
rotating disc is confined to the very central regions and a more
slowly rotating spheroid occurs beyond 1~R$_e$. 
Our class A galaxy V/$\sigma$ profiles are generally consistent
with the Wu et al. (2014) predictions for their simulated class A 
galaxies.  


The other fast rotators (classes B and D) 
are shown in Fig. 5. Major mergers are expected to
spin up the remnant galaxy (Di Matteo et al. 2009). Indeed, the
galaxies we have classified as B or D? show uniformly rapidly rising
profiles.  They exceed, even at 3~R$_e$, the Wu et al. (2014)
mean V/$\sigma$ predictions at 5~R$_e$ for classes B and D.

We find that the slow rotators 
(classes C, E and F), and class A galaxies have 
local V/$\sigma$ profiles that appear consistent with 
the expectations of their respective assembly class from Wu et al. (2014). 
In the case of class B galaxies, the simulations may under-predict their 
rotational support. We note that 
the predictions depend on the nature of the feedback 
incorporated in the models, and we remind the reader that 
the Wu et al. simulations do not include 
AGN feedback.  However, Dubois et al. (2013) have showed that AGN feedback 
tends to reduce the predicted V/$\sigma$ even further. 

Although Coccato et al. (2009) did not publish V/$\sigma$ profiles
from their PNe, they did calculate the average V/$\sigma$ for all PNe
over their field-of-view and compared this value to an average for the
starlight component in the inner regions. They classified galaxies in
one of three profile types, i.e. high and rising V/$\sigma$ profiles
(NGC 1023 and 4564), relatively flat profiles (NGC 821, 3377, 4494,
4697), low V/$\sigma$ in the inner regions and either flat or rising
values beyond (NGC 3608, 4374 and 5846). 
In section 4, we assigned NGC 821, 3377, 4494 and 4697 all to class A, and NGC
3608, 4374 and 5846 to the slow rotator classes E and F. Although NGC
1023 and NGC 4564 both have rising profiles they were assigned to
class B (alternative A) and class A (alternative B) respectively.
Thus our V/$\sigma$ profiles are
in excellent agreement with the PNe profiles of Coccato et al. (2009).

\begin{figure}
\begin{center}
\includegraphics[scale=0.43,angle=0]{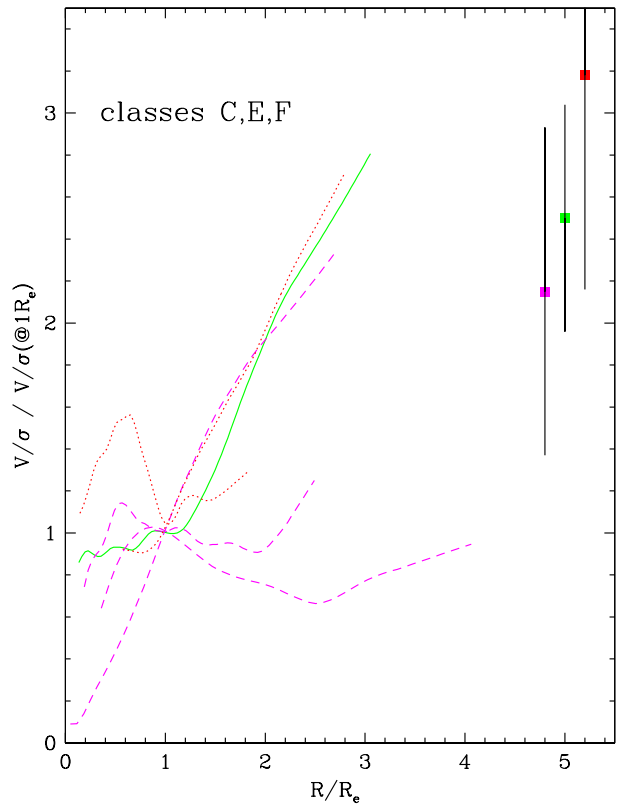}
\caption{Normalised V/$\sigma$ profiles for slow rotator SLUGGS galaxies. The profiles are normalised at 1~R$_e$ 
and are colour-coded by assembly class (i.e. C = green solid, E = magenta dashed, 
F = red dotted). 
The mean values and the error on the mean for the simulated slow rotators (i.e. classes C, E and F) from 
Wu et al. (2014) at 5~R$_e$ are shown on the right side of the plot, and slightly offset for clarity.
}
\end{center}
\end{figure}

\begin{figure}
\begin{center}
\includegraphics[scale=0.43,angle=0]{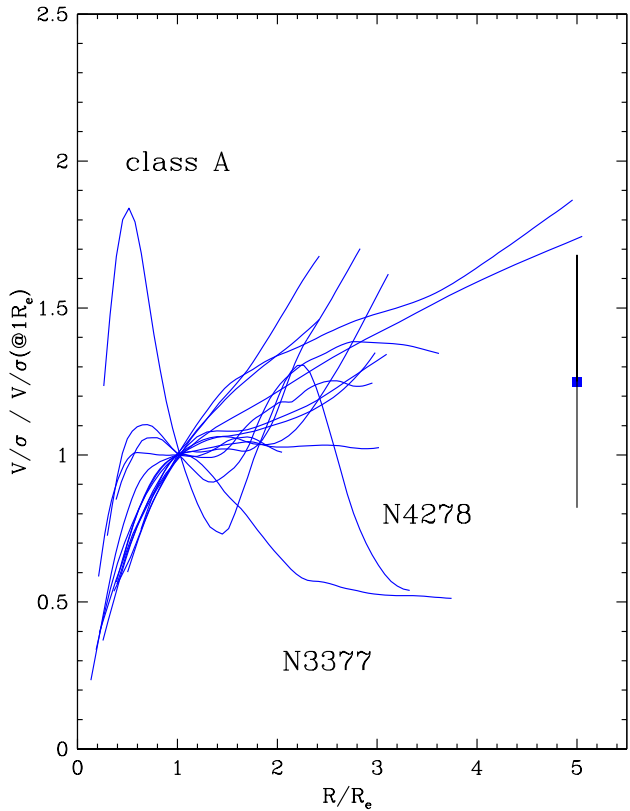}
\caption{Normalised V/$\sigma$ profiles for class A SLUGGS galaxies. The profiles are 
normalised at 1~R$_e$, and are coloured blue solid for assembly class A. 
The mean value and the error on the mean for the simulated class A galaxies from 
Wu et al. (2014) at 5~R$_e$ is shown on the right side of the plot. We highlight two galaxies whose profiles peak at small radii and then decline to lower values at large radii 
(i.e. NGC 3377 and NGC 4278). 
}
\end{center}
\end{figure}

\begin{figure}
\begin{center}
\includegraphics[scale=0.43,angle=0]{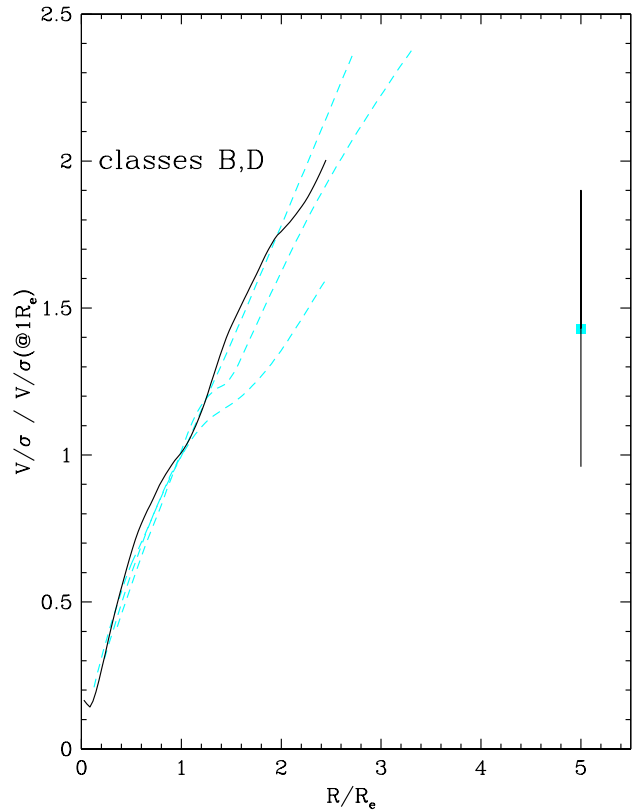}
\caption{Normalised V/$\sigma$ profiles for fast rotator 
classes B and D SLUGGS galaxies. The profiles are normalised at 1~R$_e$, and are coloured cyan dashed for assembly class B and black solid for class D.
The mean value and the error on the mean for the simulated class B galaxies from 
Wu et al. (2014) at 5~R$_e$ is shown on the right side of the plot. 
}
\end{center}
\end{figure}

\section{Trends with Ex-situ fraction}

The fraction of stars formed externally and later accreted in minor or
major mergers is predicted to correlate with the mean age of the
stellar population, the total mass density slope and the metallicity
gradient. Below we investigate whether measurements of these
properties for our sample reveal a trend with the mean ex-situ
fraction for the assembly class that we have assigned to each galaxy. We
also examine the trend with galaxy stellar mass, which is expected to
show a weak dependence on ex-situ fraction. An important caveat to note is that 
the ex-situ fractions that we use from N14 
are likely to be over-estimated due to the lack of realistic 
feedback (e.g. see Dubois et al. 2013 for the effects of incorporating AGN feedback).

\subsection{Mean stellar ages}

The N14 simulations predict a strong dependence on the fraction of ex-situ
formed stars with galaxy age, i.e. the galaxies that have
accreted a larger fraction of their stars (classes D, E and F) tend to
have, on average, older stars. These classes have experienced dry mergers with less gas
dissipation than classes A, B and C. Although classes A, B and C 
have lower ex-situ fractions, they have nonetheless 
formed some new stars in recent 
gas-rich mergers and, hence, they tend to
display younger average ages.

Stellar ages averaged over all radii for each SLUGGS galaxy are not
available but as a proxy we use the mass-weighted stellar age within 1
R$_e$ as measured by McDermid et al. (2015). We supplement this with
luminosity-weighted ages for NGC 3115 (9.0 Gyr) from Norris et al. (2006), and 
NGC 1400 (13.8 Gyr) and NGC
1407 (12.0 Gyr) from Spolaor et al. (2008).

In Fig. 6 we plot the stellar age of SLUGGS galaxies against their
stellar mass and their inferred fraction of stars formed externally
and later accreted (i.e. ex-situ fraction).  The trend of age with
galaxy mass is similar to those seen in larger samples (e.g. Denicolo
et al. 2005; McDermid et al. 2015) with high mass galaxies being very
old, and lower mass galaxies showing a large range of ages. When coded
by assembly class (we follow the same colour scheme as N14, except
yellow is replaced by black for the print version of the journal), the
plot shows that class A galaxies tend to be of lower mass. 
young/intermediate aged galaxies are exclusively of class A. 
location of the data points in this plot do not depend on the assembly
class that we have assigned to SLUGGS galaxies in this work.

N14 did not list the stellar ages for 
individual galaxies, only a mean value for each class. In Fig. 6 we show 
the mean values of the zoom simulations for the six assembly classes. 
We find that the 6 SLUGGS galaxies of classes D, E and F, 
with high ex-situ fractions, are ``consistently old'' as
predicted by N14. Galaxies of classes A, B and C, with lower ex-situ
fractions, are well described by the statement in N14 that classes A, B
and C ``are younger and show a larger spread in age
(some have ages similar to classes D, E and F, some are as young as
$\sim$ 8.5 Gyr).''
The young/intermediate aged SLUGGS galaxies are exclusively of class A. 
There appear to be no 
SLUGGS galaxies that contain a significant young stellar population as 
the result of recent gas-rich merger.
The average ages of the simulated galaxies are 
generally younger than the observed mass-weighted ages of the SLUGGS 
galaxies. The introduction of AGN feedback into the models, and the associated suppression of 
in-situ star formation, would be expected to lead to systematically older model ages (and lower stellar masses). 
Overall the assembly classes for the SLUGGS galaxies are
consistent with the relative trends of stellar mass and ex-situ fraction with age 
from the N14 simulations..

\begin{figure}
\begin{center}
\includegraphics[scale=0.43,angle=0]{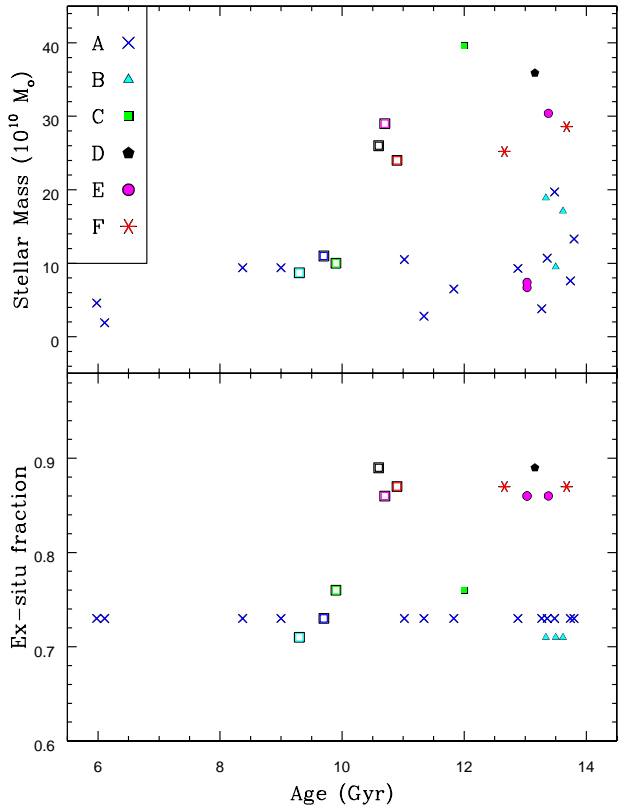}
\caption{Mean stellar age versus galaxy stellar mass and fraction of 
ex-situ formed stars. Symbols, as per the legend, show the 
mass-weighted stellar ages within 1 R$_e$ for SLUGGS galaxies from 
McDermid et al. (2015), and luminosity-weighted ages for 
NGC 1400 and 1407 from Spoloar et al. (2008). 
The open square symbols show the mean values for the 
six assembly classes from the simulations of N14.
Symbols are colour-coded according to their assembly class.
(i.e. A = blue, B = cyan, C = green, D = black, E = magenta, F = red).  
In the lower panel, the SLUGGS galaxies are 
assigned mean ex-situ fractions based on the average value for their assigned 
assembly class. 
}
\end{center}
\end{figure}

\subsection{Total mass density slopes}

Remus et al. (2013) investigated the total mass (i.e. stellar plus
dark matter) density profile for a subset of the N14 models that were re-simulated at 
two different spatial resolutions. 
The slope of this profile ($\gamma_{tot}$) has been
measured observationally in massive early-type galaxies using strong lensing 
(Auger et al. 2010), weak lensing (Gavazzi et al. 2007) 
and from the combination of ATLAS$^{\rm 3D}$ and SLUGGS 
2D kinematics (Cappellari et al. 2015). These studies all found $\gamma_{tot}$ 
to be close to -2 (i.e. an isothermal profile) over a wide range of 
galactocentric radii. Remus et al. examined, among 
other quantities, how $\gamma_{tot}$ depends on the host galaxy 
stellar mass and on the fraction of ex-situ formed stars.

In Fig.  7 we show the total mass density slope of the 
simulated galaxies from Remus et al. versus their stellar mass and ex-situ 
fraction since z = 2 taken from table 1 of N14. 
A clear trend is seen for the simulated galaxies 
in both plots with more massive galaxies, and those 
that have accreted a larger fraction of their total mass, having slightly 
shallower total mass density profiles. 

\begin{figure}
\begin{center}
\includegraphics[scale=0.43,angle=0]{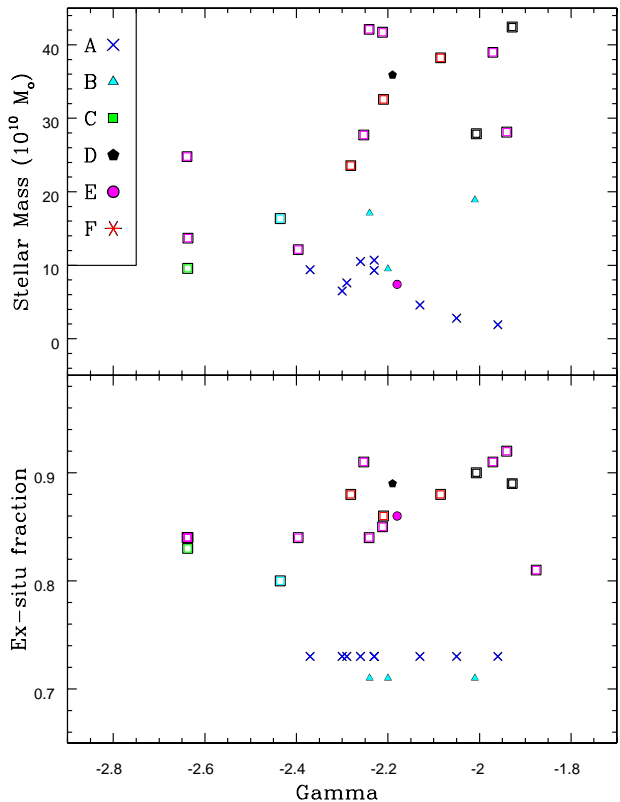}
\caption{Total mass density slope ($\gamma_{tot}$) versus galaxy stellar 
mass and fraction of 
ex-situ formed stars. 
Symbols, as per the legend, show the SLUGGS galaxies from Cappellari et al. (2015). 
The open square symbols show the cosmological zoom 
simulated galaxies of Remus et al. (2013).
Symbols are colour-coded according to their assembly class.
(i.e. A = blue, B = cyan, C = green, D = black, E = magenta, F = red).  
In the lower panel, the SLUGGS galaxies are 
assigned mean ex-situ fractions based on the average value for their assigned 
assembly class. 
 }
\end{center}
\end{figure}

In the top panel we overplot the $\gamma_{tot}$ 
results from Cappellari et al. (2015)
for 14 fast, regular rotators. Our galaxies tend to be of lower mass
on average than the simulated ones, and there is no clear indication of a 
stellar mass trend with $\gamma_{tot}$.  

In the lower panel we overplot the SLUGGS galaxies from this work,  again using 
$\gamma_{tot}$ from Cappellari et al. and the mean ex-situ fraction of stars 
associated with each assembly class.  N14 uses a fiducial value of 
0.18 in-situ fraction, or 0.82 ex-situ fraction, to differentiate the assembly 
classes of A, B and C from D, E and F. The simulated galaxies of 
Remus et al. are mostly the slow rotator classes E and F with high 
ex-situ fractions. In contrast, we have classified most SLUGGS 
galaxies to be of class A or B which have low ex-situ fractions 
on average. 
There is no clear trend for the ex-situ fraction to vary with 
total mass density slope for SLUGGS galaxies. 

It is difficult to draw strong conclusions given the small samples of
both observed and simulated galaxies, and that they probe different
mass ranges, assembly histories and ex-situ fractions.  However, this initial
investigation suggests that the simulations do not yet capture the
degree to which the total mass density slope is independent of the present day
galaxy mass and its assembly history. This suggests some missing physics.
Indeed, the addition of feedback from AGN and winds appears to weaken the
dependence of $\gamma_{tot}$ on stellar mass (Remus et al., in prep.).   
With larger samples of observed galaxies,
one can also test the prediction that galaxies dominated by minor merger
mass growth (i.e. classes A and F) should experience halo expansion
and lower central dark matter densities (Johansson et al. 2009).

\subsection{Metallicity gradients}

Stellar metallicity gradients for a subsample of massive galaxies from the zoom
simulations of Hirschmann et al. (2013),  which are 
based on the same initial conditions as N14, 
were in investigated by Hirschmann et al. (2015).
Despite not including AGN feedback, they found that feedback in the
form of momentum-driven winds provided a good match to the colour
profiles of nearby galaxies. The predicted metallicity profiles fell
steeply in the inner regions and flattened out to a near constant
metallicity at large radii. This feedback has the effect of increasing
the in-situ fractions, and hence stellar masses, relative to the
original N14 simulations. In Figure 8 we plot the Hirschmann et
al. (2015) values for the ex-situ fraction and stellar mass versus
their metallicity gradients between 2 and 6 R$_e$. 
flat to a gradient of about -1 dex per dex.  As shown in Figure 7, The
metallicity gradient correlates with the relative fraction of accreted
stars, so that galaxies with a large ex-situ fraction have mildly
shallower gradients on average. The simulated galaxies are
colour-coded by their N14 assembly class. 
The introduction of feedback significantly
reduces the fraction of ex-situ formed stars, and affects the merger history 
(e.g. reducing the number of minor mergers.)

The stellar metallicity gradients of SLUGGS galaxies 
have been measured by Pastorello et
al. (2014) and are updated here using new data with an improved method
for measuring the gradients and their errors (see Appendix B). Inner
(0.32 to 1 R$_e$) and outer (1 to 2.5 R$_e$) gradients are
measured. It is not possible to compare to stellar metallicity gradients
to 6 R$_e$ as done in the simulations of Hirschmann et al. (but see
Pastorello et al. 2015 for globular cluster metallicity gradients
to comparable effective radii). The gradients measured for the SLUGGS galaxies 
to
2.5 R$_e$ are generally much steeper than the values quoted in
Hirschmann et al. (2015) as they are measuring gradients at much larger radii
where the metallicity 
profiles have flattened off to a relatively constant metallicity. 
Nevertheless, in Fig. 8 we
compare the outer gradients of the SLUGGS galaxies (Table B1) with
those from the Wind Model (WM) of present day galaxies by Hirschmann et al.

Fig. 8 shows that both datasets reveal a weak trend of metallicity
gradient with stellar mass, i.e. steeper gradients tend to be found in
lower mass galaxies. 
A similar trend was found by Spolaor et
al. (2010) for the inner $<$1 R$_e$ metallicity gradients in a sample
of early-type galaxies.  
The steepest gradients tend to be found in class A SLUGGS galaxies. 
We also show the metallicity gradient against
ex-situ fraction. For the SLUGGS galaxies, the ex-situ fraction is the
mean value corresponding to the assigned assembly class. The plot indicates
that the galaxies with a high fraction of ex-situ formed stars tend to
have relatively shallow metallicity gradients. Galaxies with lower
ex-situ fractions reveal a large range of metallicity gradient.

\begin{figure}
\begin{center}
\includegraphics[scale=0.43,angle=0]{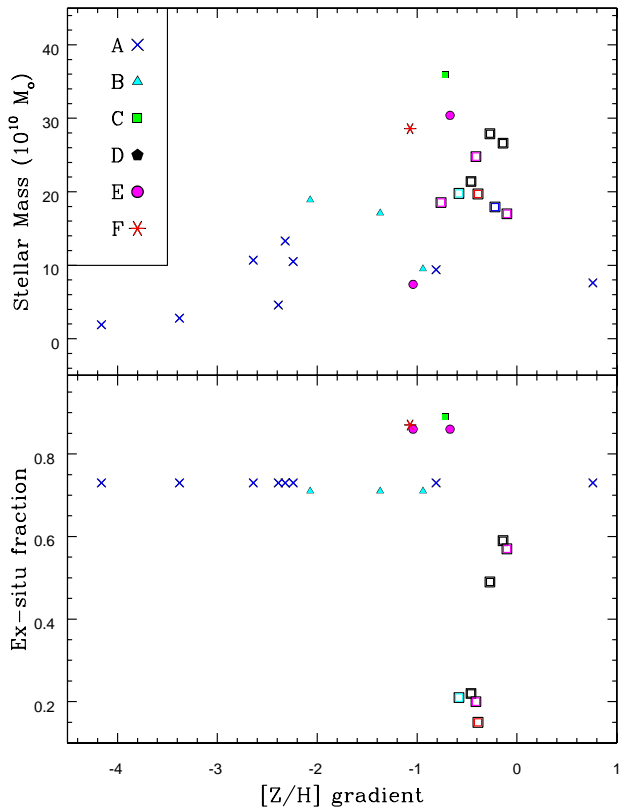}
\caption{Stellar metallicity gradient versus galaxy stellar mass and fraction of 
ex-situ formed stars. Symbols, as per the legend, show the SLUGGS galaxies 
with gradients measured from 1 to 2.5 R$_e$. 
The open square symbols show the cosmological zoom 
simulated galaxies of Hirschmann et al. (2015) with feedback from momentum-driven  
winds. The stellar masses, ex-situ fractions and 
gradients (measured from 2 to 6 R$_e$) are from 
Hirschmann et al. (2015). 
Symbols are colour-coded according to their N14 assembly class.
(i.e. A = blue, B = cyan, C = green, D = black, E = magenta, F = red).  
Both the simulations and SLUGGS galaxies show a 
weak trend for steeper gradients to be found in lower mass galaxies, 
and high ex-situ fractions are 
associated with galaxies that have relatively shallow gradients. 
}
\end{center}
\end{figure}

\section{Discussion}

We find that the bulk of the SLUGGS galaxies (14 out of 24)
can be classified as assembly class A.  These are centrally fast
rotators with an embedded disc, which often gives way to a slower
rotating spheroid component.  Three galaxies are classified as class B,
which also contain discs but continue to rotate rapidly at large
radii. One galaxy is tentatively assigned to fast rotator class D.  
The remaining six galaxies are all slow rotators (classes C, E and F). 

Before discussing our results further 
we present a summary plot in Fig. 9 which  shows  the distribution of assembly classes as a function of stellar mass and $\lambda_R$. Here $\lambda_R$ is the 
average value measured within 1 R$_e$ from Emsellem et al. (2011) and supplemented by Arnold et al. (2014) for those 
SLUGGS galaxies not in the ATLAS$^{\rm 3D}$ survey. It is a proxy for the angular momentum in galaxies, separating centrally fast and slow rotators. Figure 9 can be 
compared with figure 6 from N14, which uses a similar colour scheme for the different assembly classes, 
bearing in mind that both samples have small numbers.

As with N14, we find that low mass galaxies with high central rotation
tend to be of class A or B, and higher mass galaxies tend to be slowly
rotating ones of class E and F. However, some interesting differences
are also present.  We find many more class A galaxies (58\%) compared
to the frequency in N14 (21\%). Furthermore, class A is also more
common than class B, particularly for the fast central rotators. 
If our small sample of 24 galaxies is representative of 
normal $\sim$10$^{11}$ M${_\odot}$ 
early-type galaxies in the local Universe, then most galaxies
have grown in mass and size by the accretion of stars and gas (and globular clusters)  
from smaller galaxies, with major mergers playing a lesser role.

In the N14 simulations most class C galaxies are of low mass. We have
none, with the only tentative class C galaxy being the high mass NGC
1407 (and alternative class of D). Similarly, in the simulations 50\% of the galaxies have
experienced gas-poor mergers (i.e. classes D, E and F.). This is in stark
contrast to the SLUGGS galaxies with only 25\% assigned to these 3 dry
merger assembly classes.

\begin{figure}
\begin{center}
\includegraphics[scale=0.43,angle=0]{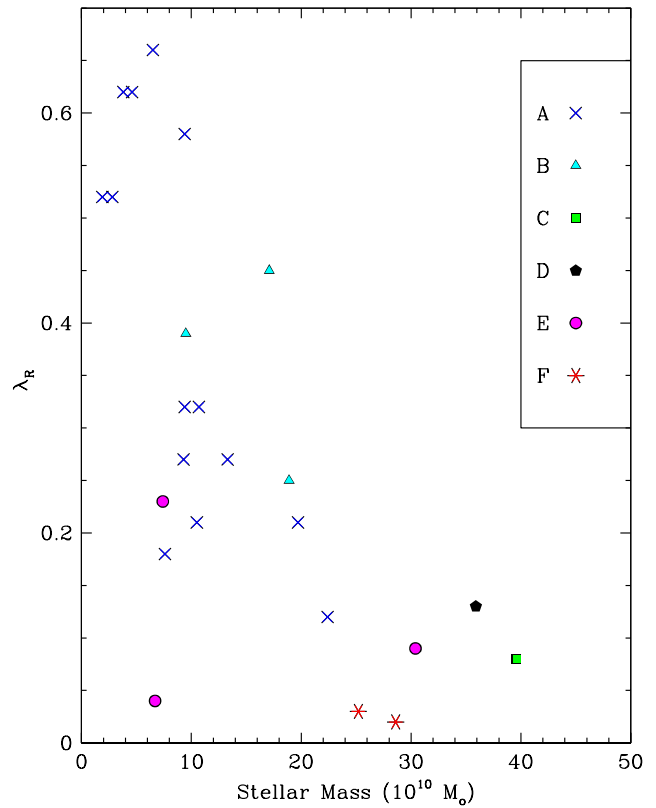}
\caption{Galaxy stellar mass as a function of central angular momentum. The $\lambda_R$ 
parameter is a proxy for the angular momentum within 1 R$_e$ (with data from 
the ATLAS$^{\rm 3D}$ survey by Emsellem et al. 2011, supplemented with SLUGGS data by 
Arnold et al. 2014). Symbols, as per the legend,  
show the SLUGGS galaxies colour-coded according to their 
assembly class, i.e. A = blue, B = cyan, C = green, D = black, 
E = magenta, F = red. 
This plot can be compared to figure 6 of Naab et al. (2014) 
which shows the hydrodynamical zoom simulations.
The low mass galaxies with high central rotation tend to be of class A or B, 
and higher mass galaxies tend to be slowly rotating ones of class C, E or F.
}
\end{center}
\end{figure}

The main caveat about the relative frequency of the different assembly
classes noted by N14 is that their sample is small and therefore not
statistically representative of the galaxy population. However,
another key limitation is the correct handling of feedback which
affects essentially all of the properties discussed in this paper at
some level. It is thus useful to compare our results with the
observations of the ATLAS$^{\rm 3D}$ survey of 260 early-type
galaxies. We find that 75\% of our galaxies are classified as fast
rotators (i.e. classes A, B and D), which is comparable to the 86\% of
the ATLAS$^{\rm 3D}$ survey classified as centrally fast rotators
(Emsellem et al. 2011).

As noted in the Introduction, we are only aware of one independent study  
that has previously assigned an N14 assembly class, i.e. class A to NGC 4697 by Spiniello 
et al. 2015). In this work, we find it shows clear characteristics of class A. And we also note that 
their age for the bulge component (13.5 $\pm$ 1.4 Gyr) agrees well with the age derived by McDermid et al. (2015) of 13.4 Gyr.  
Here we tentatively assign the same class D to NGC 4649 as Pota et al. 
(2015) using SLUGGS data. 

With over half of our sample assigned to class A it appears that the
most frequent assembly pathway for early-type galaxies of mass $\sim$
10$^{11}$ M$_{\odot}$ is one of many gas-rich minor mergers since z =
2 (i.e. in the last 10 Gyr) and a significant fraction of in-situ star
formation. 
These minor mergers tend to reduce the angular
momentum in the outer regions so that the $\lambda_R$ and V/$\sigma$
profiles may peak for the central disc and then decline in the halo.
We note that most class A galaxies have very old ($>$ 10 Gyr)
mass-weighted stellar ages, suggesting that the gas content of the
minor mergers did not lead to significant star formation.

Three galaxies (NGC 1023, 2768 and 4526), which have all been 
classified as S0 galaxies,  have been assigned to
class B. These galaxies have central disc-like kinematics and
similarly high in-situ fractions (0.29) as class A, however the mass
growth is dominated by gas-rich {\it major} mergers. These mergers
tend to transfer angular momentum to the halo, spinning-up the final
remnant which results in rising radial $\lambda_R$ and V/$\sigma$
profiles. Again, the gas content has not resulted in significant
recent star formation as all four galaxies contain very old stars.

We tentatively assign class D to NGC 4649 (M60), with class B as the alternative. It is a 
fast rotator like class B galaxies, but the evidence for a central kinematic 
disk is weaker. Class D, as the result of a gas-poor major merger, is in line with 
its old stellar age that indicates no recent central star formation. 

Three galaxies (NGC 3608, 4365 and 4473) have been assigned class
E. The mass growth of class E galaxies is expected to be dominated by
recent gas-poor major mergers, although minor mergers also contribute.
They should contain predominately old stars with little, or no,
younger stars.  This class of slow rotators includes simulated
galaxies that reveal kinematic twists, minor axis rotation, KDCs and
the rare 2-$\sigma$ galaxies. Indeed, NGC 3608 and NGC 4365 are known
to harbour KDCs, and NGC 4473 is one of only 4\% of galaxies in the
ATLAS$^{\rm 3D}$ sample to reveal a double peak in the 2D velocity
dispersion map (i.e. a 2-$\sigma$ galaxy).  All three have old
mass-weighted ages (McDermid et al. 2015).  The kinematics of the
stars and globular clusters in NGC 4473 reveal mild rotation in the
very central regions, which gives way to a complex largely
non-rotating halo (Foster et al. 2013; Alabi et al. 2015).  NGC 4365
reveals minor axis rotation (i.e. the galaxy rolls rather than spins)
beyond its central KDC region (Arnold et al. 2014). The globular
cluster system of NGC 4365 indicates an ongoing interaction with a
smaller galaxy in the group (Blom et al. 2012).  NGC 3608 is a rare
example of a non-regular rotator in the ATLAS$^{\rm 3D}$ survey
(Krajnovic et al. 2011). We note that the simulations of 
Hoffman et al. (2009, 2010) suggested an alternative means of forming central 
KDCs, i.e. gas-rich binary mergers. However, both NGC 3608 and NGC 4365 have 
very old mean stellar ages ($\ge$ 13 Gyr) along with h$_3$ and h$_4$ vs 
V/$\sigma$ distributions (see Figures A5 and A7) that resemble class 
E (gas-poor mergers) rather than the distributions predicted by Hoffman et 
al. (2009).

NGC 5846 is a very slowly rotating massive galaxy at the centre of its
group. As is common with other slow rotators, it is found in a high
density environment (Fogarty et al. 2014).  
We tentatively assign it class F?. We also
assign class F to NGC 4374 (M84).  The mass growth of such galaxies is
almost exclusively via the accretion of gas-poor minor mergers with
very little in-situ star formation.  The repeated minor mergers
decrease the angular momentum of the galaxy over time.  Galaxies of
this class tend to be very round and host an old stellar population --
NGC 5846 has an ellipticity of 0.08 and mass weighted stellar age of
12.7 Gyr, while NGC 4374 has ellipticity 0.05 and age 13.7 Gyr.

NGC 1407 is also a massive galaxy at the centre of its group. Its 2D
velocity map reveals more net rotation than NGC 5846, however we
tentatively classify it as a slow rotator of class C? (and alternative class D). Its 2D velocity
dispersion map (and 2D metallicity map) shows a lot of substructure, which may be 
indicative of a gas-rich major merger (Schauer et al. 2014).
The mass growth of a class
C galaxy is dominated by a recent ($\le$ 10 Gyr ago) gas-rich major
merger. In this case, the major merger
has led to a spin-down of the remnant so that it has little angular
momentum today. We note that the galaxy has a uniformly old age of 12 Gyr
within 1 R$_e$ (Spolaor et al. 2008), suggesting that if a recent
gaseous merger has occured it has not led to significant star
formation in the galaxy inner regions.  


Simulated present
day galaxies with high fractions of externally formed stars are
expected to have older ages (N14), shallower total mass density
profiles (Remus et al. 2013) and shallower metallicity gradients
(Hirschmann et al. 2015).  Such galaxies also tend to be more massive
on average.
With the caveat that there can be quite a spread in the ex-situ
fraction within a given assembly class, which can reduce any trend
compared to expectations, we have investigated the predicted trends of
age, mass density slope and metallicity gradient with both the mean
ex-situ fraction and galaxy stellar mass. We find that the high
ex-situ fraction galaxies, in general, have old ages and shallow
metallicity gradients as predicted (Figures 6 and 8). However, our
sample galaxies do not show any evidence for shallower total mass
density slopes (Figure 7), contrary to predictions. The zoom
simulations need to probe galaxies of lower mass and include
more that have assembly pathways with lower ex-situ fractions
(i.e. classes A, B and C) before strong conclusions can be drawn.

\section{Conclusions and Future Work}

Using stellar kinematic data from the SLUGGS survey we compare the
properties of 24 massive (M$_{\ast}$ $\sim$ 10$^{11}$ M$_{\odot}$)
early-type galaxies with the predictions from the hydrodynamical zoom
simulations of 44 galaxies by Naab et al. (2014). In particular, we
assign each SLUGGS galaxy into one of six Naab et al. assembly classes
(i.e. evolutionary pathways) based on three diagnostics: the behaviour
of the higher order moments h$_3$ and h$_4$ versus V/$\sigma$, their
1D local $\lambda_R$ profile, and their 2D kinematic map. We assign an
assembly class that best represents the kinematic characteristics of
each galaxy as defined in Naab et al. We also check the normalised
V/$\sigma$ profile for each galaxy with that expected for the mean
value of its assembly class from Wu et al. (2014).

Of the 24 galaxies examined, we assign 14 to class A. This class
reveals inner regions with disc-like kinematics and outer regions that
rotate slowly, if at all. Such galaxies are thought to have
experienced gradual gas dissipation and multiple gas-rich minor
mergers in the last 8 Gyr.  Three galaxies, all S0s, are assigned to
class B. These reveal strong inner rotation which continues
into the outer halo. Gas-rich major mergers, which spin up the outer
regions, are thought to contribute significantly for this class of
galaxy. Their old stellar ages indicate that any merger-induced star
formation happened some time ago. We tentatively assign one galaxy,
NGC 1407, to class C. If correct, this slow rotator has a significant
fraction of its stars formed in-situ and experienced a gas-rich major
merger which has led to a net spin-down of the galaxy with little or no recent star formation.  We tentatively assign 
one galaxy, NGC 4649, to class D, which are fast rotators that have
experienced a gas-poor major merger. We have assigned 3 galaxies to
class E. The simulated galaxies of class E often show kinematic
twists, decoupled cores and other kinematic features as a result of
gas-poor major mergers. Two galaxies assigned to this class (NGC 3608
and NGC 4365) are known to host kinematically decoupled cores, while
the third NGC 4473 is a rare 2-$\sigma$ galaxy (showing double peaks
in its 2D velocity dispersion map). Two slow rotators are assigned to
class F (i.e. NGC 4374 and NGC 5846). Both are massive, round
ellipticals. For this class, multiple gas-poor minor mergers have
contributed significant mass growth in the last 10 Gyr and lead to a
spin-down of the final remnant galaxy. We conclude that most SLUGGS
galaxies have assembly pathways indicating mass and size growth has
occurred via the accretion of stars and gas (and globular clusters)
from minor ($<$ 1:4) rather than major ($>$ 1:4) mergers. This is in
contrast to Naab et al. (2014) in which most of their simulated galaxies have
undergone a major merger.

As predicted in the simulations, we find that the fraction of accreted
stars in the last 10 Gyr (from the mean value of each assembly class)
correlates with the observed stellar age and metallicity gradient for
SLUGGS galaxies.  However, unlike the prediction of Remus et
al. (2013), the observed total mass density slope does not appear to
correlate with the fraction of accreted stars. Confirmation of this
result requires simulations with a better match in stellar mass and
assembly histories to the SLUGGS survey galaxies.

Future work should include an analysis of the remaining 3 SLUGGS
galaxies (i.e. NGC 4459, 4474, 4486).  We also aim to examine the
V$_{rms}$ profiles of SLUGGS galaxies and to compare with them 
Remus et al. (2013).  They predicted that the radial profiles will be
very flat and largely independent of mass, environment and assembly
history.

The simulated galaxies of Hirschmann et al. (2015) reveal a correlation
of stellar metallicity gradient from 2--6 R$_e$ with the fraction of accreted stars.  
However, measuring such a radial range from integrated
starlight is extremely difficult due to the very low surface
brightness in these halo galaxy regions. A more promising observational approach is
to obtain the metallicities of large numbers of halo globular clusters (see Pastorello et al. 2015).

The spatial resolution of the Naab et al. (2014) and Wu et al. (2014)
simulations was around 0.4 kpc. This limits details of the stellar
kinematics to be greater than 0.2 R$_e$ and may restrict the ability
of the simulations to produce thin and elongated disks (as observed in
nature) and to properly resolve KDCs.  Future simulations at much
higher resolution may reveal variations in the predicted kinematic
properties of the current 6 assembly classes and perhaps even reveal
additional assembly pathways (see Moody et al. 2014). A more detailed
investigation of inclination effects on the various kinematic
parameters is also needed. Finally, a key limitation of the Naab et
al. (2014) models and related works used in this paper, is the lack of
AGN and strong stellar feedback. Such feedback needs to be
incorporated into the next generation of models.

The combination of detailed observations in the outer halo regions of
early-type galaxies with high resolution hydrodynamical simulations
holds much promise in enabling us to reconstruct the evolutionary
histories of individual galaxies.

\section{Acknowledgements}

We thank R.-S. Remus for a helpful discussion.
We thank the staff of the W. M. Keck Observatory for their
support. Some the data presented herein were obtained at the W.M. Keck
Observatory, which is operated as a scientific partnership among the
California Institute of Technology, the University of California and
the National Aeronautics and Space Administration.  DAF thanks the ARC
for financial support via DP130100388. This research was in part
supported by NSF grant AST-1211995.

\section{References}

Alabi A.~B., et al., 2015, MNRAS, 452, 2208 \\
Arnold J.~A., et al., 2014, ApJ, 791, 80 \\
Auger M.~W., Treu T., Bolton A.~S., Gavazzi R., Koopmans L.~V.~E., Marshall 
P.~J., Moustakas L.~A., Burles S., 2010, ApJ, 724, 511 \\
Blom C., Forbes D.~A., Brodie J.~P., Foster C., Romanowsky A.~J., Spitler 
L.~R., Strader J., 2012, MNRAS, 426, 1959 \\
Brodie J.~P., et al., 2014, ApJ, 796, 52\\
Cappellari M., et al., 2011, MNRAS, 413, 813 \\
Cappellari M., et al., 2015, ApJ, 804, L21 \\
Ceverino D., Dekel A., Bournaud F., 2010, MNRAS, 404, 2151\\
Coccato L., et al., 2009, MNRAS, 394, 1249 \\
Cortesi A., et al., 2013, MNRAS, 432, 1010 \\
Dekel A., Sari R., Ceverino D., 2009, ApJ, 703, 785\\
Denicol{\'o} G., Terlevich R., Terlevich 
E., Forbes D.~A., Terlevich A., 2005, MNRAS, 358, 813 \\
Di Matteo P., Pipino A., Lehnert M.~D., Combes F., Semelin B., 2009, A\&A, 499, 427 \\
Dubois Y., Gavazzi R., Peirani S., Silk J., 2013, MNRAS, 433, 3297\\
Emsellem E., et al., 2004, MNRAS, 352, 721 \\
Emsellem E., et al., 2011, MNRAS, 414, 888 \\
Fogarty L.~M.~R., et al., 2014, MNRAS, 
443, 485 \\
Foster C., et al., 2011, MNRAS, 415, 3393 \\
Foster C., Arnold J.~A., Forbes D.~A., Pastorello N., Romanowsky A.~J., 
Spitler L.~R., Strader J., Brodie J.~P., 2013, MNRAS, 435, 3587 \\
Foster, C. et al. 2015, MNRAS, submitted\\
Gavazzi R., Treu T., Rhodes J.~D., 
Koopmans L.~V.~E., Bolton A.~S., Burles S., Massey R.~J., Moustakas L.~A., 
2007, ApJ, 667, 176 \\
Hirschmann M., et al., 2013, MNRAS, 436, 2929 \\
Hirschmann M., Naab T., Ostriker J.~P., 
Forbes D.~A., Duc P.-A., Dav{\'e} R., Oser L., Karabal E., 2015, MNRAS, 
449, 528 \\
Hoffman L., Cox T.~J., Dutta S., Hernquist 
L., 2009, ApJ, 705, 920 \\
Hoffman L., Cox T.~J., Dutta S., Hernquist 
L., 2010, ApJ, 723, 818 \\
Johansson P.~H., Naab T., Ostriker J.~P., 2009, ApJ, 697, L38 \\
Kere{\v s} D., Katz N., Weinberg D.~H., Dav{\'e} R., 2005, MNRAS, 363, 2 \\
Khochfar S., et al., 2011, MNRAS, 417, 845 \\
Krajnovi{\'c} D., et al., 2011, MNRAS, 
414, 2923 \\
Lackner C.~N., Cen R., Ostriker J.~P., Joung M.~R., 2012, MNRAS, 425, 641 \\
McDermid R.~M., et al., 2015, MNRAS, 448, 
3484 \\
Moody C.~E., Romanowsky A.~J., Cox T.~J., Novak G.~S., Primack J.~R., 2014, 
MNRAS, 444, 1475 \\
Norris M.~A., Sharples R.~M., Kuntschner H., 2006, MNRAS, 367, 815 \\
Naab T., Johansson P.~H., Ostriker J.~P., Efstathiou G., 2007, ApJ, 658, 710\\ 
Naab T., et al., 2014, MNRAS, 444, 3357 \\
Oser L., Ostriker J.~P., Naab T., Johansson P.~H., Burkert A., 2010, ApJ, 
725, 2312 \\
Oser L., Naab T., Ostriker J.~P., Johansson P.~H., 2012, ApJ, 744, 63 \\
Pastorello N., Forbes D.~A., Foster C., 
Brodie J.~P., Usher C., Romanowsky A.~J., Strader J., Arnold J.~A., 2014, 
MNRAS, 442, 1003 \\
Pota, V., et al., 2015, MNRAS, 450, 3345\\
Remus R.-S., Burkert A., Dolag K., Johansson P.~H., Naab T., Oser L., 
Thomas J., 2013, ApJ, 766, 71 \\
Schauer A.~T.~P., Remus R.-S., Burkert A., 
Johansson P.~H., 2014, ApJ, 783, L32 \\
Scott N., Graham A.~W., Schombert J., 2013, ApJ, 768, 76 \\
Spiniello C., Napolitano N.~R., Coccato 
L., Pota V., Romanowsky A.~J., Tortora C., Covone G., Capaccioli M., 2015, 
MNRAS, 452, 99 \\
Spolaor M., Forbes D.~A., Proctor R.~N., 
Hau G.~K.~T., Brough S., 2008, MNRAS, 385, 675 \\
Spolaor M., Kobayashi C., Forbes D.~A., 
Couch W.~J., Hau G.~K.~T., 2010, MNRAS, 408, 272 \\
Wu X., Gerhard O., Naab T., Oser L., Martinez-Valpuesta I., Hilz M., Churazov 
E., Lyskova N., 2014, MNRAS, 438, 2701 \\
Zolotov A., et al., 2015, MNRAS, 450, 2327\\

\appendix

\section{h$_3$ and h$_4$ vs V/$\sigma$ diagnostic plots}

In Figures A1 to A12 we show h$_3$ and h$_4$ against V/$\sigma$ for
the SLUGGS galaxies. Blue filled symbols indicate slit locations that are
within 1 R$_e$ of the galaxy centre, while red open symbols show those slit
locations beyond 1 R$_e$ from SLUGGS data only. Class A galaxies have
anti-correlated h$_3$ and V-shaped h$_4$ with V/$\sigma$; class B have
anti-correlated h$_3$ and h$_4$ clustered around zero; class C and D
have both h$_3$ and h$_4$ clustered around zero; class E and F show a
large range in h$_3$ and h$_4$ values with V/$\sigma$ clustered around
zero. These Figures can be compared with figure 9 from N14.

\begin{figure}
\begin{center}
\includegraphics[scale=0.38,angle=0]{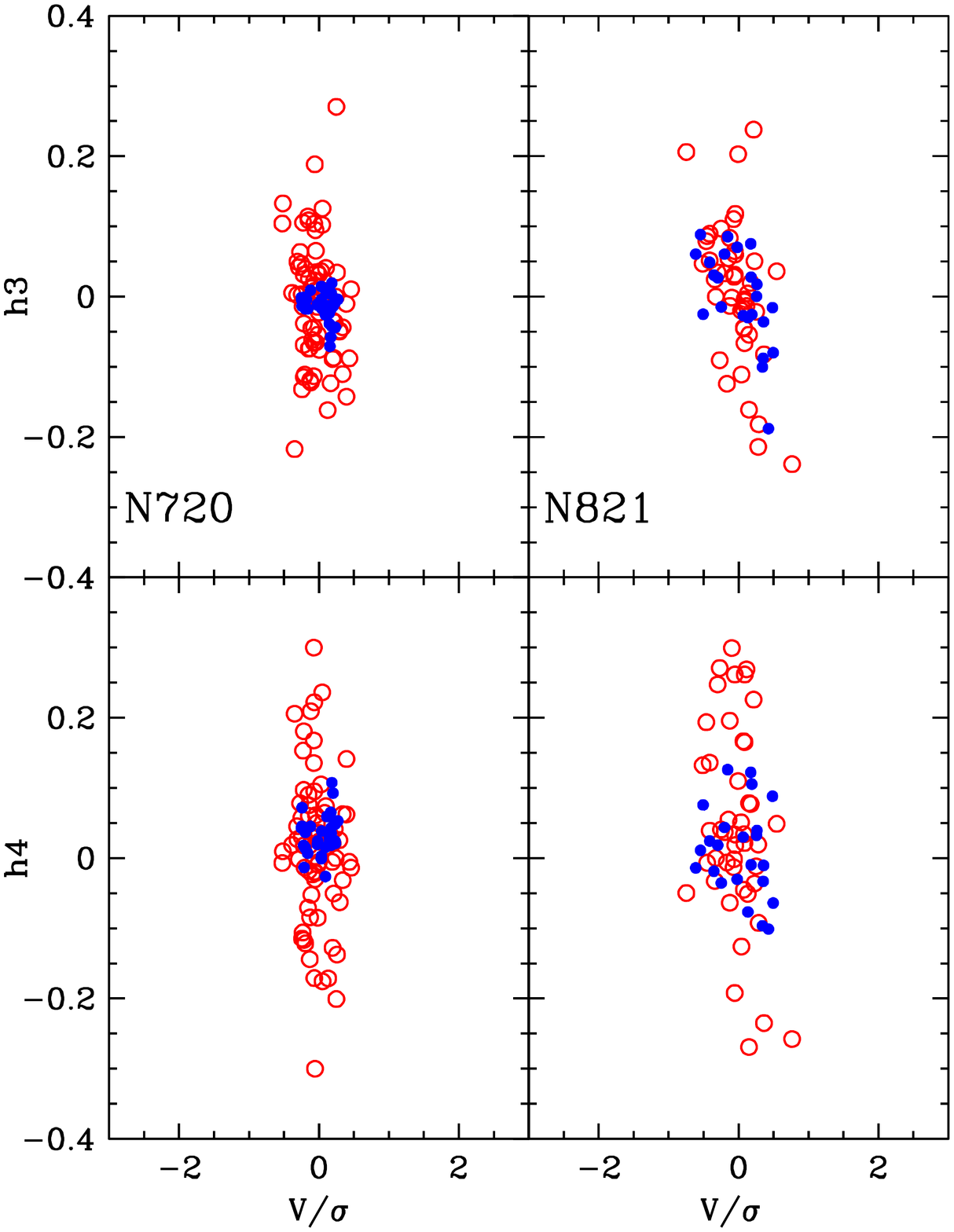}
\caption{h$_3$ and h$_4$ vs V/$\sigma$ for NGC 720 and NGC 821. Blue filled symbols 
show locations within 1 R$_e$, while red open symbols show locations beyond 
1 R$_e$. 
}
\end{center}
\end{figure}

\begin{figure}
\begin{center}
\includegraphics[scale=0.38,angle=0]{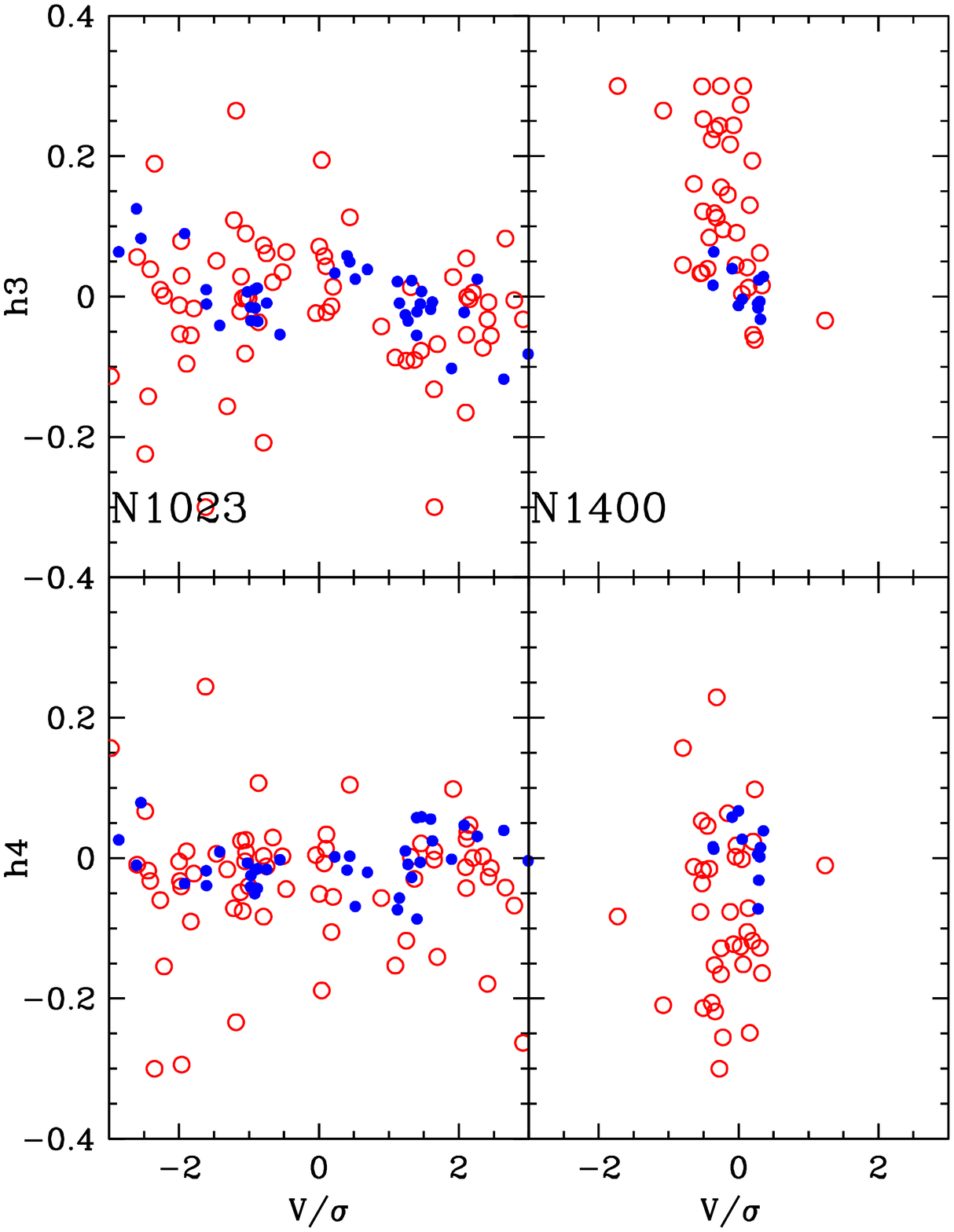}
\caption{h$_3$ and h$_4$ vs V/$\sigma$ for NGC 1023 and NGC 1400. 
}
\end{center}
\end{figure}

\begin{figure}
\begin{center}
\includegraphics[scale=0.38,angle=0]{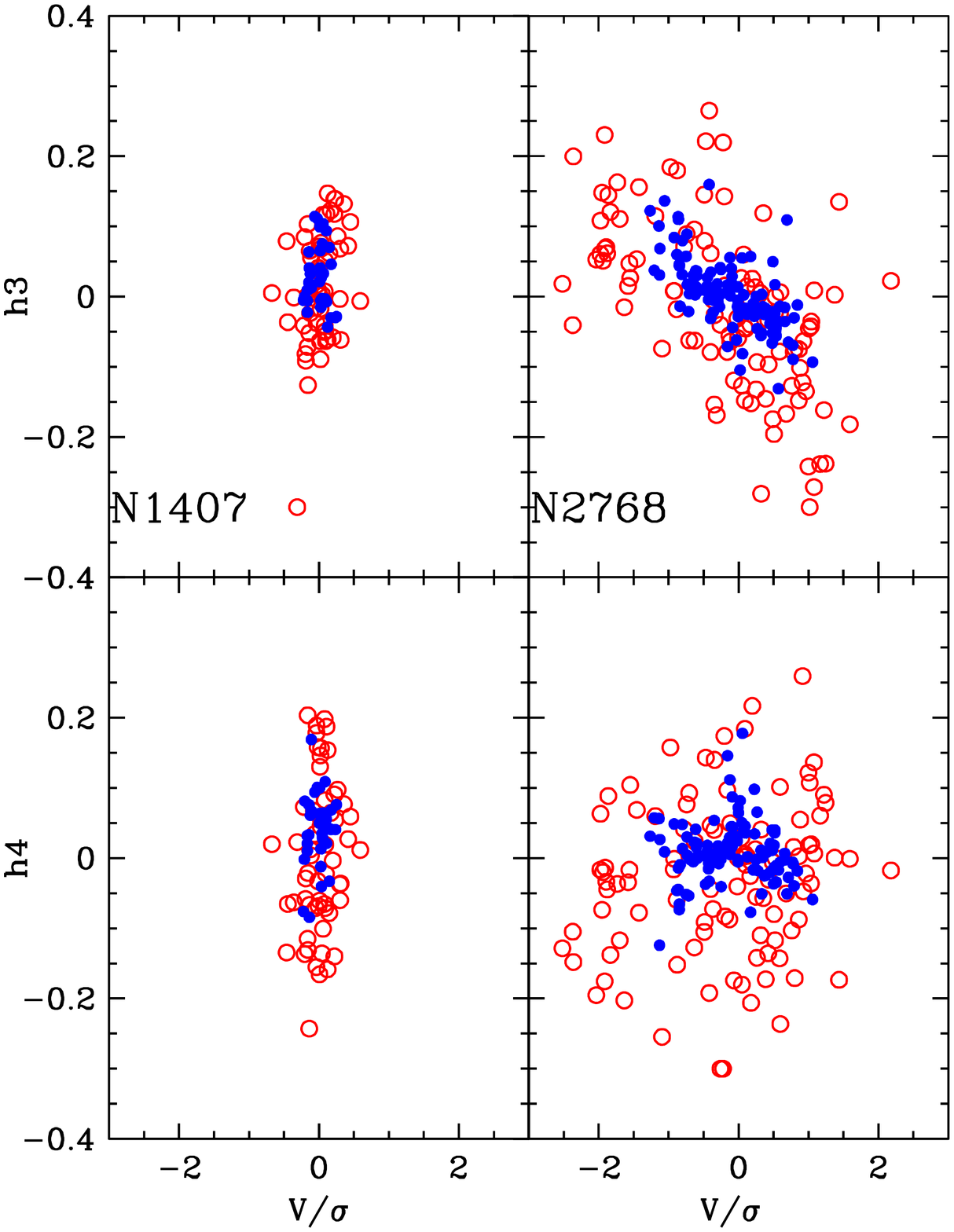}
\caption{h$_3$ and h$_4$ vs V/$\sigma$ for NGC 1407 and NGC 2768. 
}
\end{center}
\end{figure}

\begin{figure}
\begin{center}
\includegraphics[scale=0.38,angle=0]{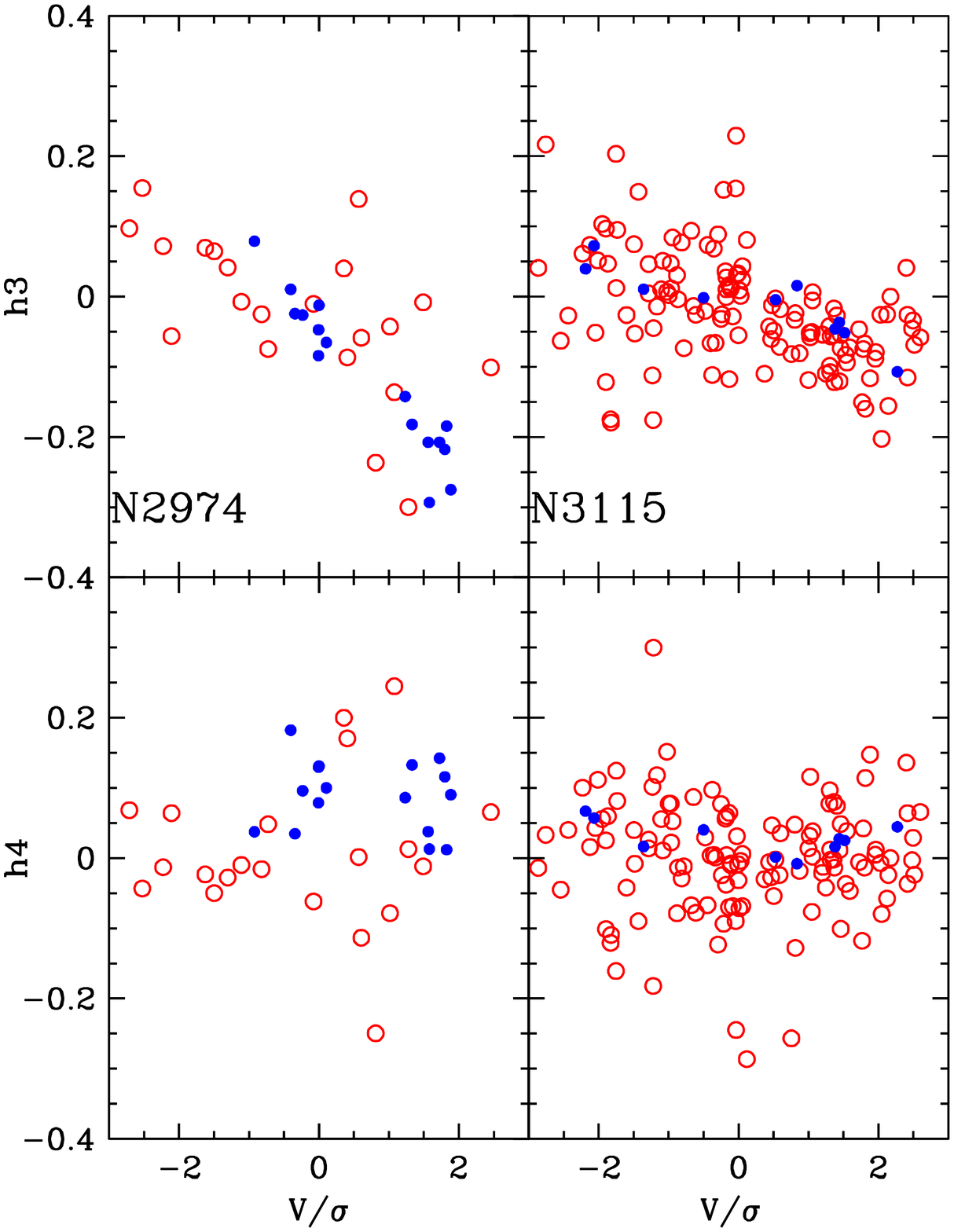}
\caption{h$_3$ and h$_4$ vs V/$\sigma$ for NGC 2974 and NGC 3115. 
}
\end{center}
\end{figure}

\begin{figure}
\begin{center}
\includegraphics[scale=0.38,angle=0]{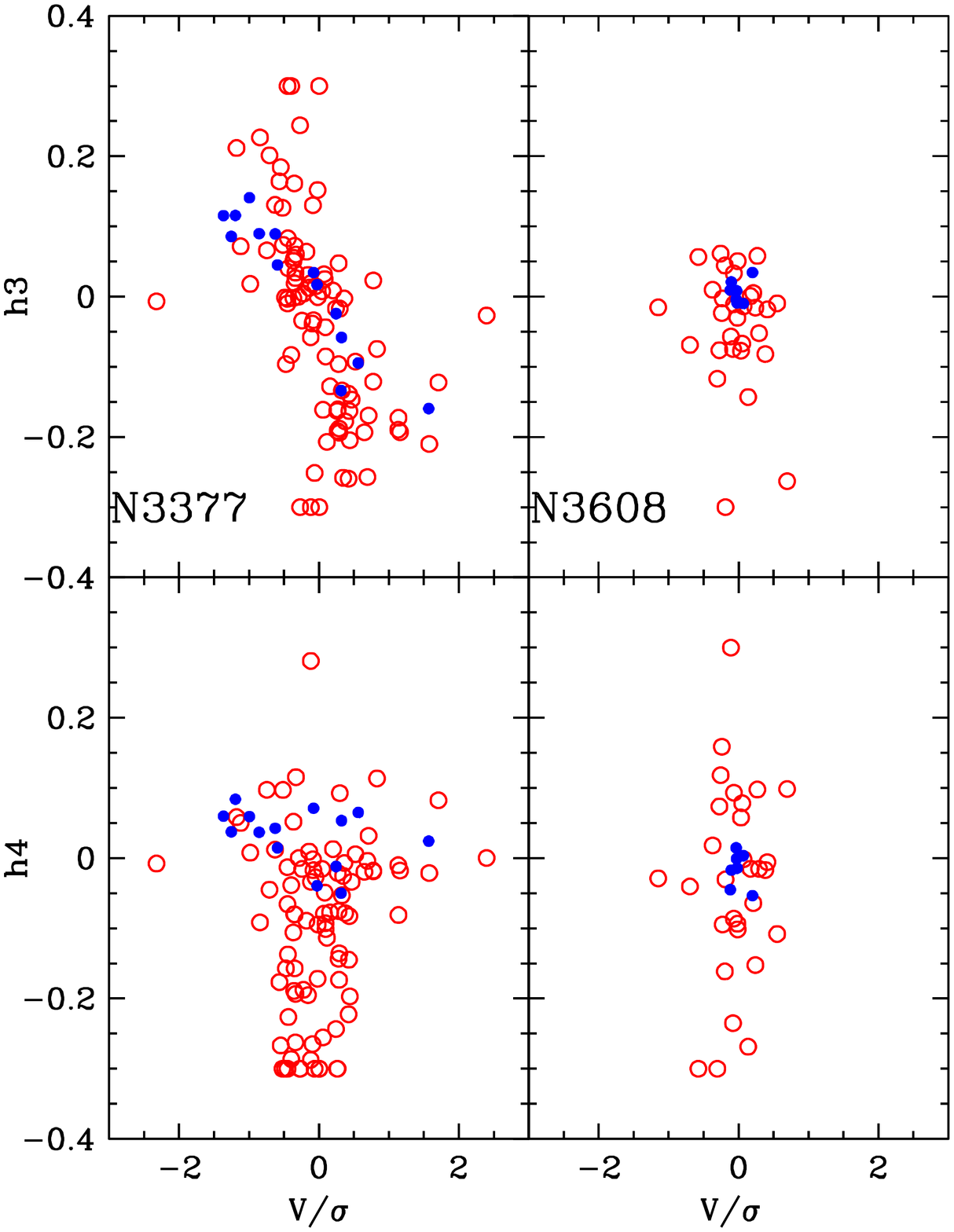}
\caption{h$_3$ and h$_4$ vs V/$\sigma$ for NGC 3377 and NGC 3608. 
}
\end{center}
\end{figure}

\begin{figure}
\begin{center}
\includegraphics[scale=0.38,angle=0]{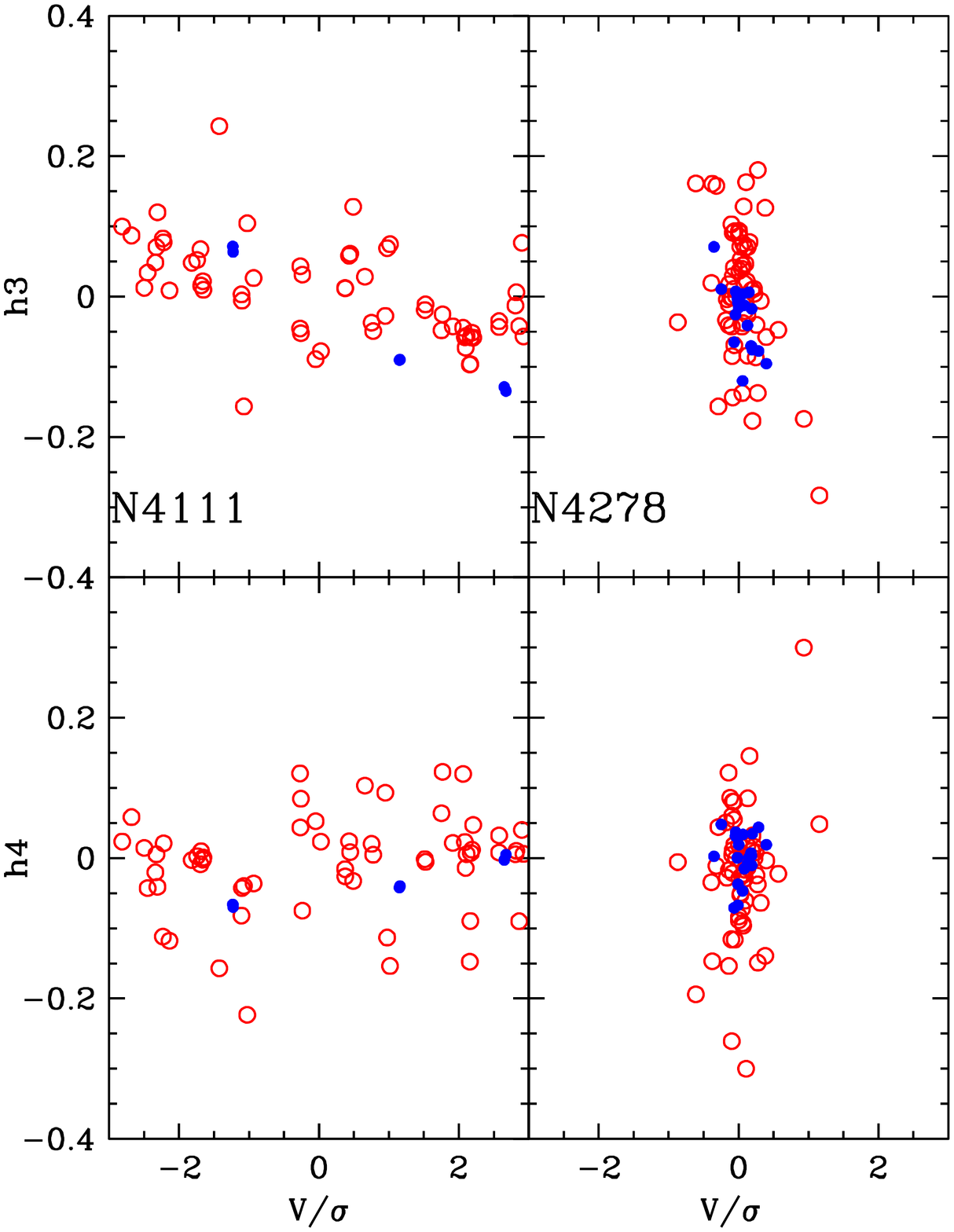}
\caption{h$_3$ and h$_4$ vs V/$\sigma$ for NGC 4111 and NGC 4278. 
}
\end{center}
\end{figure}

\begin{figure}
\begin{center}
\includegraphics[scale=0.38,angle=0]{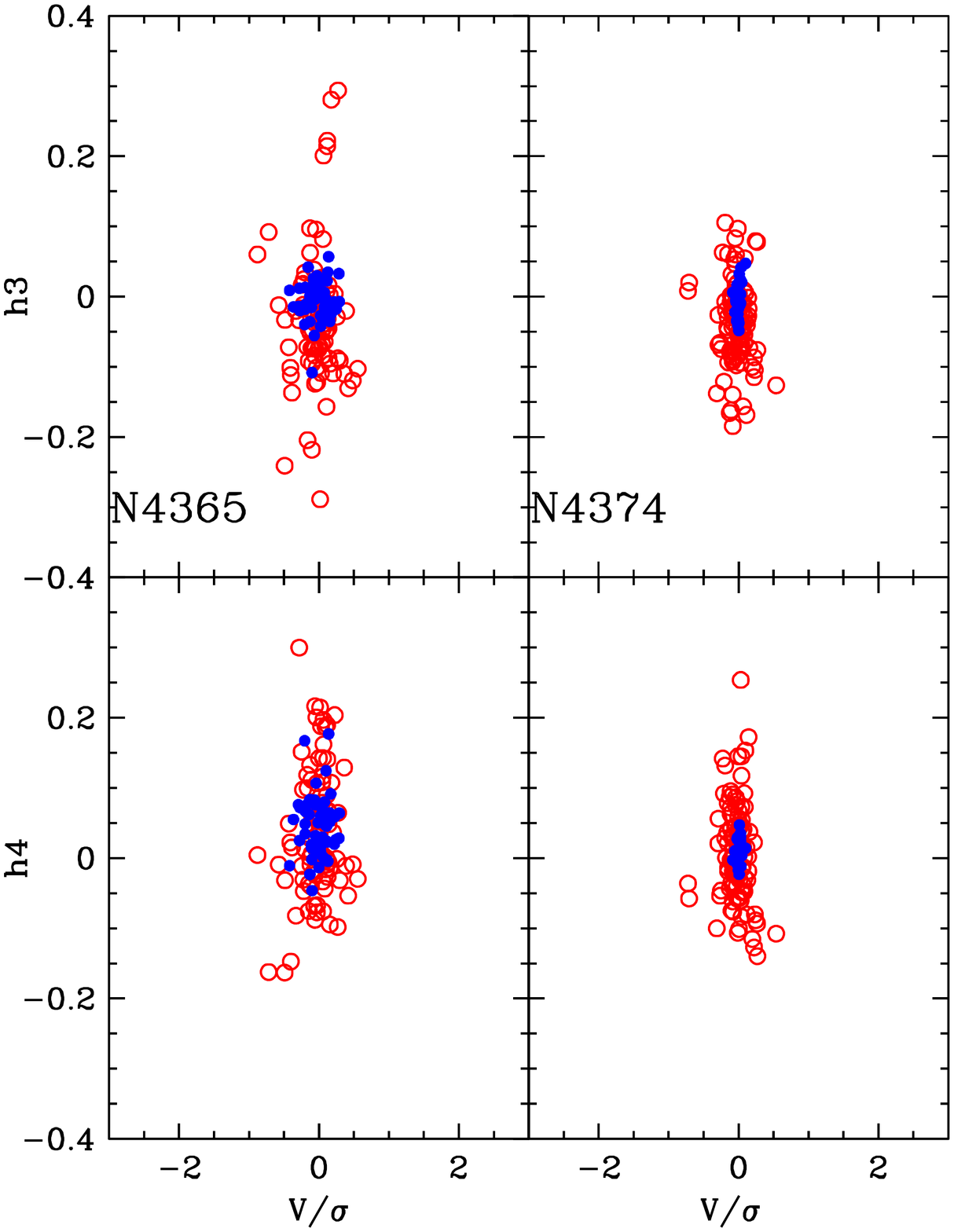}
\caption{h$_3$ and h$_4$ vs V/$\sigma$ for NGC 4365 and NGC 4374. 
}
\end{center}
\end{figure}

\begin{figure}
\begin{center}
\includegraphics[scale=0.38,angle=0]{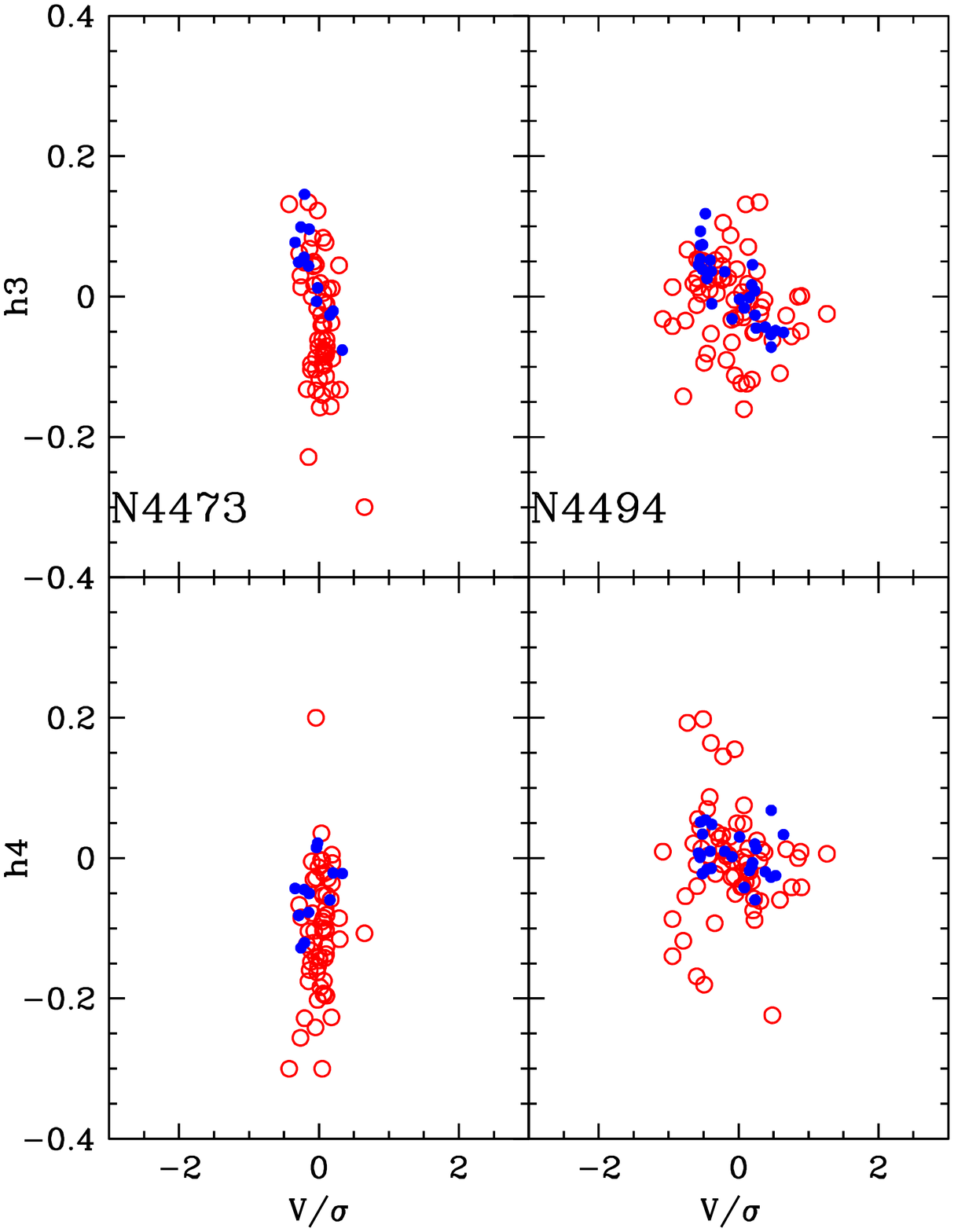}
\caption{h$_3$ and h$_4$ vs V/$\sigma$ for NGC 4473 and NGC 4494. 
}
\end{center}
\end{figure}


\begin{figure}
\begin{center}
\includegraphics[scale=0.38,angle=0]{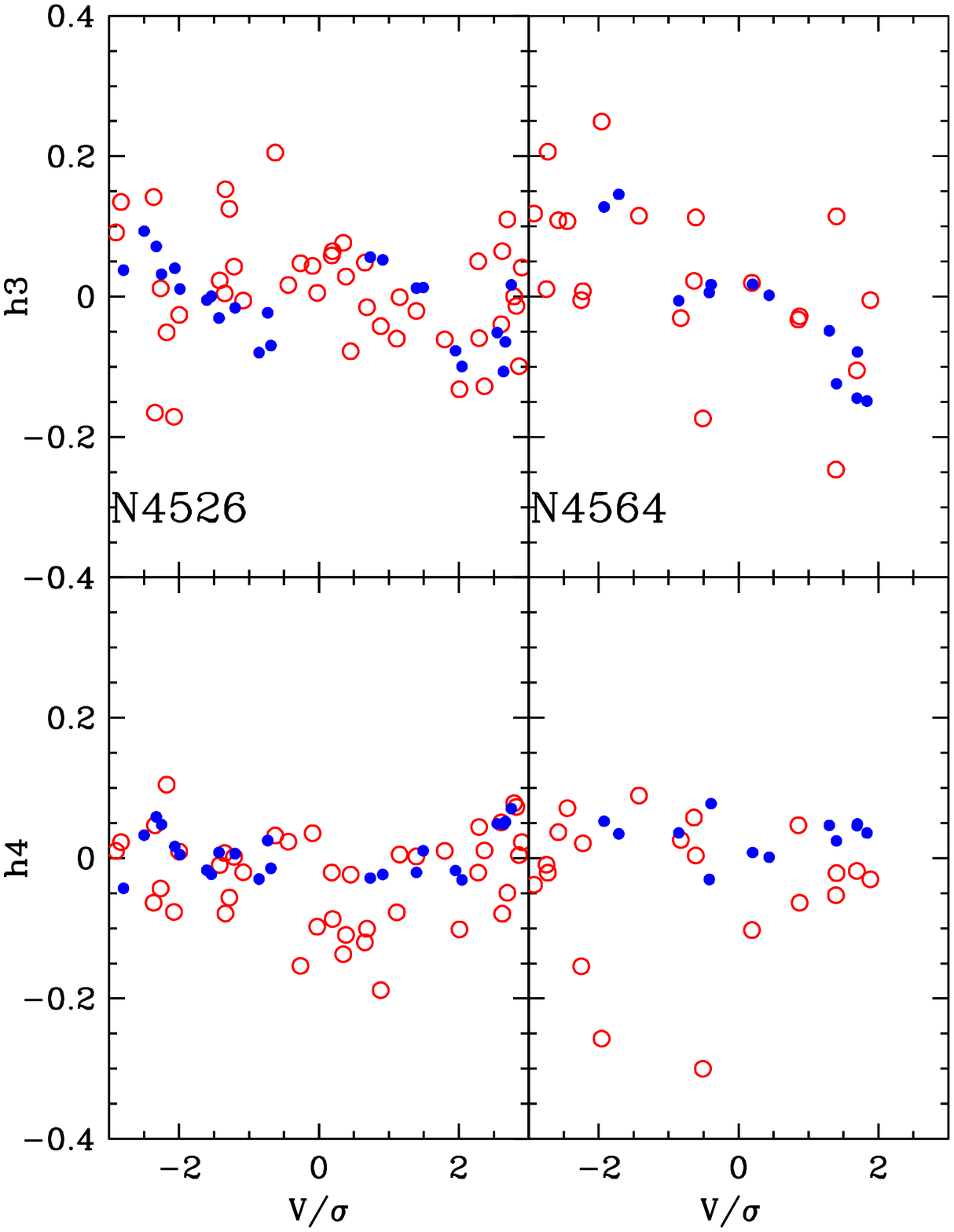}
\caption{h$_3$ and h$_4$ vs V/$\sigma$ for NGC 4526 and NGC 4564. 
}
\end{center}
\end{figure}

\begin{figure}
\begin{center}
\includegraphics[scale=0.38,angle=0]{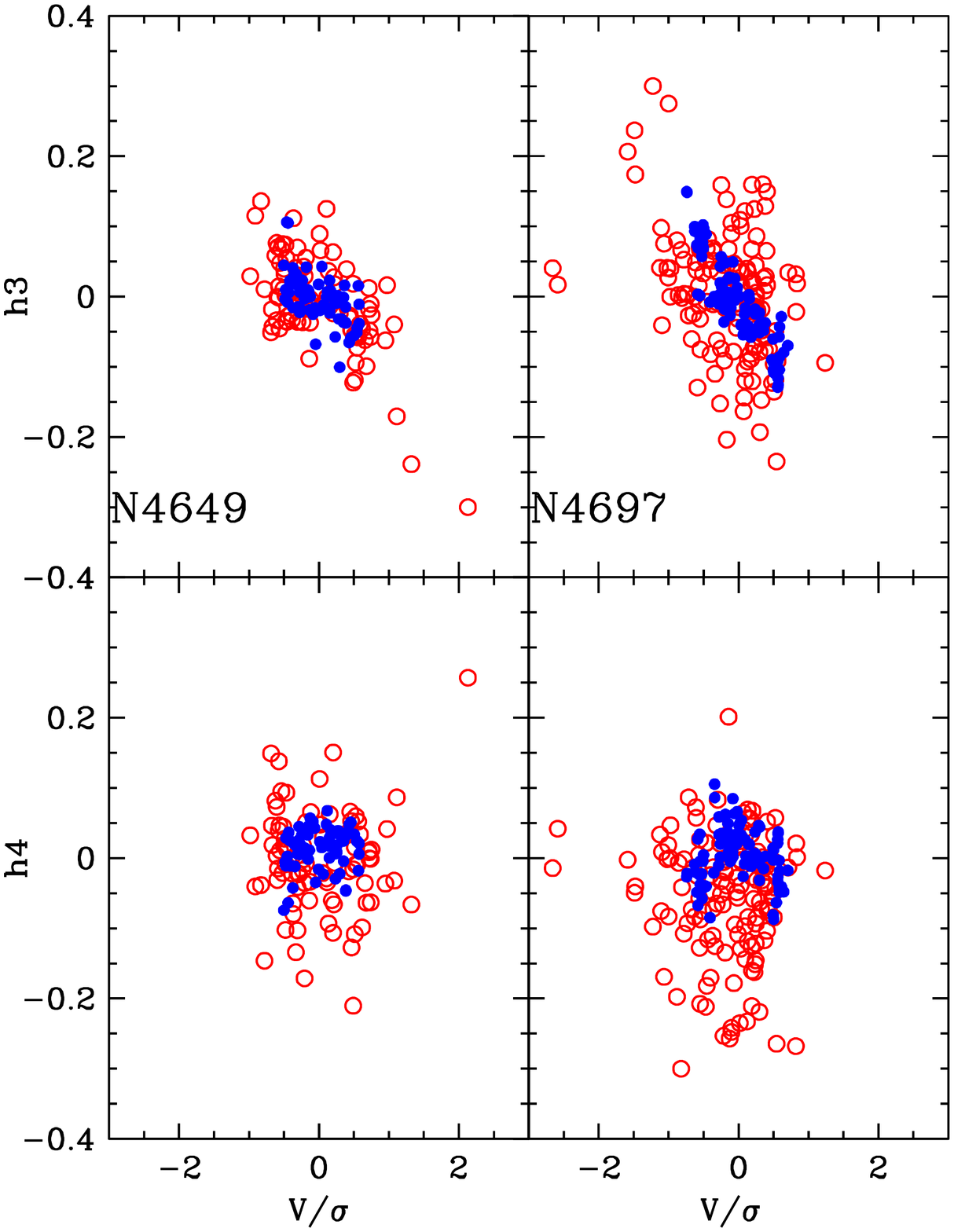}
\caption{h$_3$ and h$_4$ vs V/$\sigma$ for NGC 4649 and NGC 4697. 
}
\end{center}
\end{figure}

\begin{figure}
\begin{center}
\includegraphics[scale=0.38,angle=0]{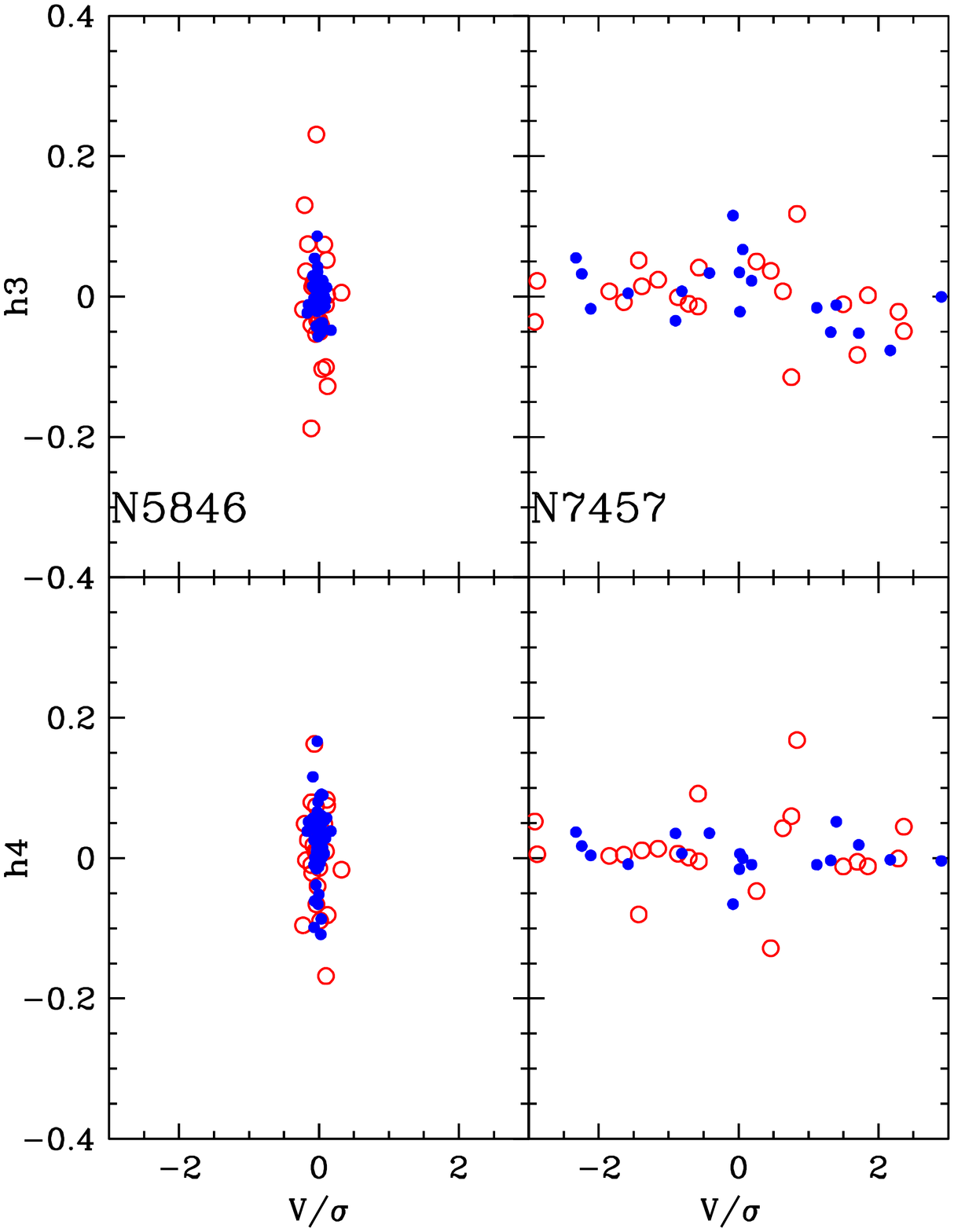}
\caption{h$_3$ and h$_4$ vs V/$\sigma$ for NGC 5846 and NGC 7457. 
}
\end{center}
\end{figure}

\begin{figure}
\begin{center}
\includegraphics[scale=0.38,angle=0]{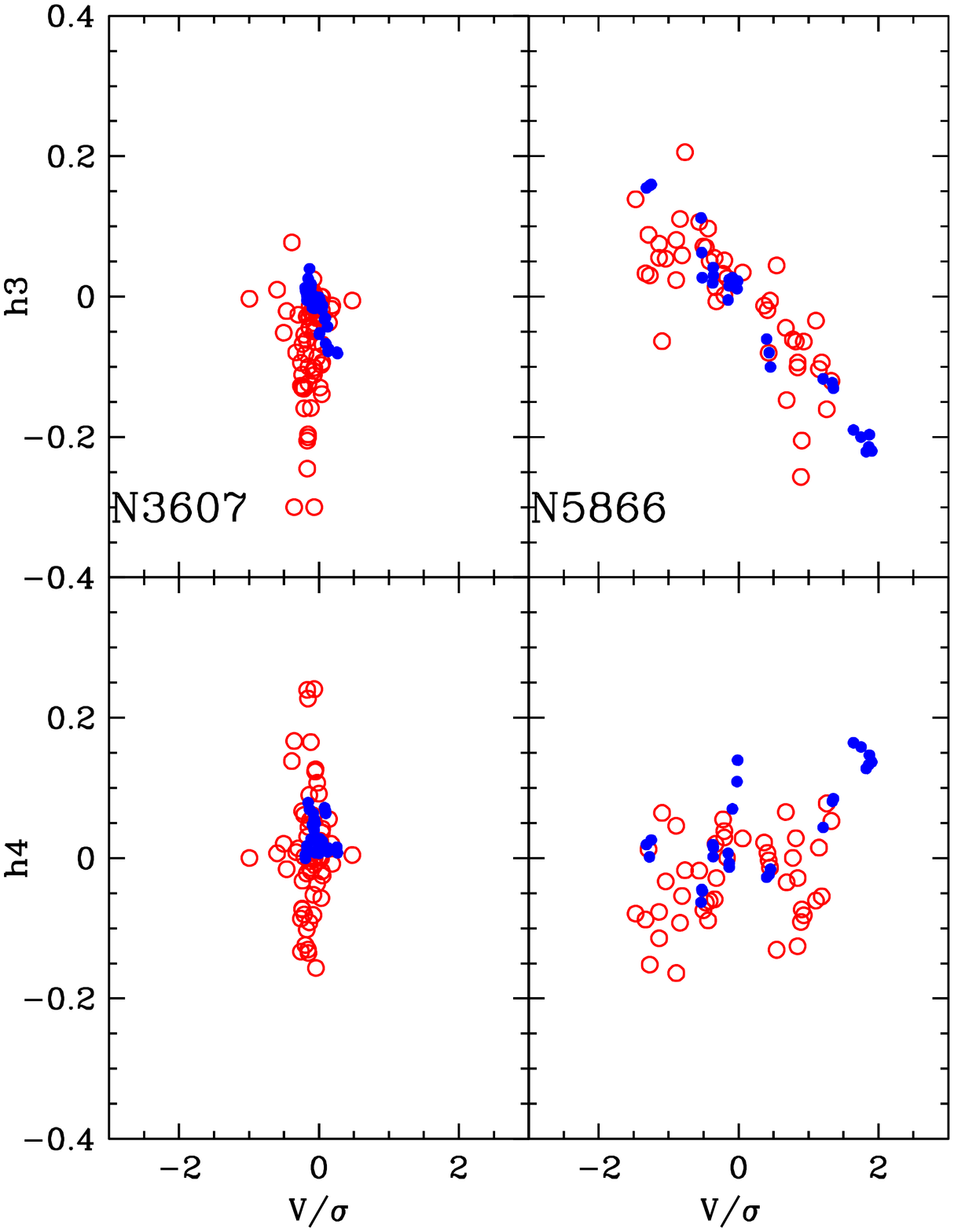}
\caption{h$_3$ and h$_4$ vs V/$\sigma$ for NGC 3607 and NGC 5866. 
}
\end{center}
\end{figure}

\section{SLUGGS galaxies metallicity gradients}

In Pastorello et al. (2014) we presented stellar metallicity 2D maps
and radial profiles for 22 SLUGGS galaxies. Since then we have obtained new
data for NGC 2768 and NGC 3115. Here we take the opportunity to
update our analysis methodology and remeasure metallicity gradients.

The initial datasets from which we obtain the 2D metallicity maps are
now cleaned from outliers in a more homogeneous fashion.  Firstly, we
remove duplicates (datapoints observed multiple times), averaging
their metallicity values by weighting on the relative errors.  Secondly,
we exclude all the points with $S/N<35$, errors on 
[Z/H] $>$ $\pm$ 0.5 dex, CaT index $<$  2 and CaT index $>$ 9. 
Thirdly, we remove all the points at galactocentric
radii $R>5~\rm{R_{e}}$, since they are likely not associated with the
galaxies under study.  Finally, all the remaining data points are
checked by eye.  This process allows us to retrieve an outer metallicity gradient 
for one  additional galaxy (NGC 4494) but excludes one galaxy (NGC 3607) 
listed in Pastorello et al. (2014). 

Another important change concerns the kriging parameters used to make the 2D maps. 
In particular, we now fit the semivariogram of all the galaxies with a third degree polynomial 
using ranges that better match the spatial sampling, i.e. 
the range$=10~\rm{arcsec}$ for all the galaxies except NGC~5846 (range $=15~\rm{arcsec}$) and NGC~7457 (range $=25~\rm{arcsec}$). 
This results in smoother 2D maps. 

The radial metallicity profiles are extracted from the kriged 2D maps
as in Pastorello et al. (2014).  However, the metallicity gradients
extracted from the radial profiles are now measured in a more robust
way.  The metallicity radial profile uncertainties are obtained with
both Monte Carlo simulations and bootstrapping.  In the former case,
we build 1000 kriged 2D maps from datasets with the same number of
data points and the same spatial position, but different associated
metallicity values.  These metallicities are randomly extracted from a
two-sided Gaussian distribution in order to obtain positive and negative
metallicity uncertainties for each measured datapoint.  In the latter
case, we also build 1000 kriged 2D maps from datasets obtained by
replacement sampling of the original dataset.  From these approaches we
generate 1D profiles and measure the standard deviation of the
metallicity values in each radial bin.  In this way we obtain separate
Monte Carlo and boostrapping confidence limits for the observed radial
metallicity profile, which we combine in quadrature to obtain the
final estimate of the total uncertainty.  We obtain a 1D metallicity
profile from each kriged 2D map.  From these sets of profiles, we
obtain a median value and a distribution of metallicity values in each
radial bin from which we measure the standard deviation as the
metallicity uncertainty.




The final updated gradients are now obtained by fitting the
metallicity profiles in equally spaced logarithmic bins with weights
that are inversely proportional to both the uncertainty and the number
of distributions from which such
uncertainties are obtained.  In Table B1 we list the updated inner
(i.e. $0.32<R<1 {R_{e}}$) and outer
(i.e. $1<R<2.5 {R_{e}}$) metallicity gradients and their
uncertainties.

The main conclusion of Pastorello et al. (2014) that outer 
metallicity gradients are steeper in lower mass galaxies is unchanged.

\begin{table}
\caption{SLUGGS galaxy stellar metallicity gradients}
\begin{tabular}{lcc}
\hline
NGC & inner [Z/H] grad. & outer [Z/H] grad.\\
 (1) & (2) & (3) \\
\hline
    NGC~1023    & -0.28$^{+0.12}_{-0.13}$  & -0.94$^{+0.49}_{-0.65}$ \\
    NGC~1400    &  -0.50$^{+0.21}_{-0.26}$  & -2.32$^{+0.62}_{-0.32}$ \\
    NGC~2768    &  -0.64$^{+0.35}_{-0.02}$  & -2.07$^{+1.66}_{-2.23}$ \\
    NGC~3115    & -0.09$^{+0.30}_{-0.32}$  & -0.81$^{+0.38}_{-0.19}$ \\
    NGC~3377    &  -0.87$^{+0.39}_{-0.33}$  & -3.38$^{+0.89}_{-0.58}$ \\
    NGC~4111    &  -0.48$^{+0.19}_{-0.26}$  & -2.39$^{+0.87}_{-0.05}$ \\
    NGC~4278    &  -0.07$^{+0.07}_{-0.09}$  & 0.76$^{+0.41}_{-0.66}$ \\
    NGC~4365    &  -0.08$^{+0.20}_{-0.14}$  & -0.67$^{+1.17}_{-1.24}$ \\
    NGC~4374    &  -0.42$^{+0.79}_{-0.63}$  & -1.07$^{+0.84}_{-1.11}$ \\
    NGC~4473    &  -0.24$^{+0.13}_{-0.34}$  & -1.04$^{+0.50}_{-0.59}$ \\
    NGC~4494    &  -0.19$^{+0.16}_{-0.25}$  & -2.24$^{+0.97}_{-0.68}$ \\
    NGC~4526    &  -0.42$^{+0.11}_{-0.07}$  & -1.37$^{+0.21}_{-0.35}$ \\
    NGC~4649    &  -0.27$^{+0.18}_{-0.07}$  & -0.72$^{+0.14}_{-0.36}$ \\
    NGC~4697    &  -0.66$^{+0.15}_{-0.03}$  & -2.64$^{+1.05}_{-0.59}$ \\
    NGC~5846    &  -0.68$^{+0.36}_{-0.03}$  & --\\
    NGC~7457    &  -0.95$^{+0.20}_{-0.26}$  & -4.16$^{+1.11}_{-2.59}$ \\
\hline

\end{tabular}
\\
Notes:    The inner metallicity gradients (second column) are measured in the radial region $0.32 < R < 1~R_{e}$, while the outer metallicity gradients (third column) are measured in the radial region $1 < R < 2.5~R_{e}$. In the cases where the 2D maps do not extend to $2.5~\rm{R_{e}}$, the radial profile is extrapolated from the available points outside $1~\rm{R_{e}}$. 
\end{table}

\end{document}